\documentclass[10pt,USenglish]{article}
\pdfoutput=1

\usepackage[latin9]{inputenc}
\usepackage{geometry}
\usepackage[USenglish]{babel}
\usepackage{verbatim}
\usepackage{prettyref}
\usepackage{amsmath}
\usepackage[normalem]{ulem}
\usepackage{amssymb}
\usepackage{graphicx}
\usepackage{setspace}
\usepackage{esint}
\usepackage{cite}
\usepackage{dsfont}

\usepackage{color}
\definecolor{darkgreen}{rgb}{0,0.5,0}
\definecolor{darkblue}{rgb}{0,0,0.6}
\definecolor{purple}{rgb}{0.4,.2,0.7}
\definecolor{awesome}{rgb}{1.0, 0.13, 0.32}

\usepackage[colorlinks=true,citecolor=darkgreen,linkcolor=purple,urlcolor=purple]{hyperref}

\numberwithin{equation}{section}
\numberwithin{table}{section}

\begin{document}
\begin{flushright}
{\tt MIT-CTP-4731}
\end{flushright}

~
\vskip15mm

\begin{center} {\Large \textsc{SUSY in Silico: numerical D-brane bound state spectroscopy}}

\vskip15mm

Tarek Anous

\vskip5mm


\it{Center for Theoretical Physics, Massachusetts Institute of Technology, Cambridge, MA 02139, USA}

\vskip5mm

\tt{ tanous@mit.edu}

\end{center}

\vskip15mm

\begin{abstract}
We numerically construct the BPS and non-BPS wavefunctions of an $\mathcal{N}=4$ quiver quantum mechanics with two Abelian nodes and a single arrow. This model captures the dynamics of a pair of wrapped D-branes interacting via a single light string mode. A dimensionless parameter $\nu$, which is inversely proportional to the Fayet-Iliopoulos parameter, controls whether the bulk of the wavefunctions are supported on the Higgs branch or the Coulomb branch. We demonstrate how the BPS and excited states morph as $\nu$ is tuned. We also numerically compute the energy gap between the ground state and the first excited states as a function of $\nu$. An expression for the gap, computed on the Coulomb branch, matches nicely with our numerics at large $\nu$ but deviates at small $\nu$ where the Higgs branch becomes the relevant description of the physics. In the appendix, we provide the Schr\"{o}dinger equations fully reduced via symmetries which, in principle, allow for the numerical determination of the entire spectrum at any point in moduli space. For the ground states, this numerical determination of the spectrum can be thought of as the first \emph{in silico} check of various Witten index calculations.

\end{abstract}


\pagebreak
\pagestyle{plain}

\setcounter{tocdepth}{3}

\tableofcontents

\section{Introduction}
The low-energy effective theory describing the dynamics of a collection of wrapped D-branes at small separations is given by quiver quantum mechanics \cite{Denef:2002ru}.  These D-branes form bound states and, depending on parameters, can either be well separated or on top of each other. When well separated, the theory is on the Coulomb branch, parametrized by a macroscopic vev $\langle|\vec{\mathbf{x}}|\rangle$ for the fields describing the separation between the branes. Conversely, when the branes are on top of each other, the theory is on the Higgs branch, parametrized by a large vev $\langle|\phi|\rangle$ for the field representing the light stretched string mode between the branes.

In this paper we will consider the simplest example of quiver quantum mechanics, describing the dynamics of a pair of D-branes with a single light stretched string mode between them. Being a supersymmetric theory, the relative separation between the branes lives in a vector multiplet $(A,\,\vec{\mathbf{x}},\,\lambda_\alpha,\,\bar{\lambda}^\beta,D)$ and the lightest stretched string mode lives in a chiral multiplet $(\phi,\,\psi_\alpha, F)$. These models have proven fruitful playgrounds for the study of supersymmetric black hole bound states at weak coupling \cite{Douglas:1996sw,Denef:2002ru,Denef:2007vg,Bena:2012hf,Lee:2012sc,Manschot:2012rx,Manschot:2013sya}. However, one obvious question remains unsolved: what is the structure of the bound state wavefunctions? While supersymmetry allows us to make some statements about the structure of the ground states, the full ground state wavefunctions remain unknown \cite{Sethi:1997pa,Sethi:2000ba}. This is because, while simpler than the full Schr\"{o}dinger equation, the BPS equations remain too difficult to solve analytically. The situation is worse for the excited states, where we have to abandon the crutch of supersymmetry altogether.

It is our interest in this paper to construct the BPS and excited states of this model. We will not be able to do so analytically, but there exist numerical methods to compute the eigenspectra of differential operators on a finite domain, see e.g. \cite{SpectralMatlab}. 
Using these techniques we numerically solve the Schr\"{o}dinger equation, plot bound state wavefunctions, and determine their energies.
In our analysis, we find that the physics is governed by a dimensionless quantity $\nu$ defined in (\ref{nu}), which is inversely proportional to the Fayet-Iliopoulos parameter $\theta$. The quantity $\nu$ dictates whether the Higgs branch or the Coulomb branch is the dominant description of the dynamics. In studying the wavefunctions and their dependence on $\nu$, we uncover that the two branches never quite decouple and the wavefunction has nonzero support in both branches for any value of $\nu$. 

Since this model is supersymmetric, much work has gone into studying features of the ground states. For example, the ground state wavefunction was previously studied in a Born-Oppenheimer approximation \cite{Smilga:1986tv,Smilga:1986rb,Smilga:2002mra}, which involves splitting the Hamiltonian $H$ into a vector part $H^v$ and a chiral part $H^c$ and putting the chiral degrees of freedom in the harmonic oscillator ground state of $H^c$. One then `integrates out' the chiral multiplet and solves the effective supersymmetric quantum mechanics on the vector multiplet degrees of freedom. This approximation is self-consistent if we are on the Coulomb branch, that is if the ground state is such that the branes are well separated, and is only valid for large values of $\vec{\mathbf{x}}$. One easy way to see this is that integrating out the chiral degrees of freedom generates a nontrivial metric on the moduli space of the vector multiplet, which appears in the effective Lagrangian as
\begin{equation}
L_{\rm kin}^{\rm eff}=\frac{1}{2}\left(\mu+\frac{1}{4\,|\vec{\mathbf{x}}|^3}\right)\dot{\vec{\mathbf{x}}}^2~.
\end{equation}
The moduli space metric is singular at the origin where the approximation breaks down.\footnote{Similar tube-like metrics arise on the Coulomb branches of the D0-D4 \cite{Sethi:2000ba} and the D1-D5 systems \cite{Douglas:1997vu}. Singularities on the Coulomb branch moduli-space usually indicate that the wavefunction spreads onto the Higgs branch near the singularity. However, the Higgs and Coulomb branches of the D1-D5 system are completely decoupled as a consequence of symmetry \cite{Witten:1995zh}, rendering the tube singularity rather enigmatic in the context of the D1-D5 CFT. The interpretation proposed in \cite{Witten:1997yu} is that the theory flows to a new CFT at the origin where $[\vec{\mathbf{x}}]$ obtains a non-trivial conformal dimension. While the Higgs and Coulomb branches do not decouple for the quiver theory of interest in this paper, whenever the quiver theory has a nontrivial superpotential then the Higgs and Coulomb branch theories appear to be described by distinct CFTs \cite{Anninos:2013nra,Anninos:2016szt,Bena:2012hf}, as occurs for D1-D5. It would be interesting to study such quivers numerically.} The structure of the ground state wavefunction near the origin has previously been out of reach, and herein we fill this gap.

Interestingly the effective quantum mechanics on the Coulomb branch of this model has an enhanced symmetry, allowing for a full determination of the spectrum on the Coulomb branch \cite{D'Hoker:1985et,D'Hoker:1985kb,Vinet:1985zh,D'Hoker:1986uh,D'Hoker:1987gr,D'Hoker:1990zy,Avery:2007xf}. In computing excited state wavefunctions we numerically compute the energy gap between the supersymmetric ground state and the first excited state as a function of $\nu$. We verify that the gap approaches the analytically determined value on the Coulomb branch as we increase $\nu$.

Recently \cite{Cordova:2014oxa,Hori:2014tda,Ohta:2014ria,Kim:2015oxa} the Witten index $W_I$, twisted by a combination of global symmetries, has been computed for various quivers; and for the model of interest in this paper, $W_I=+1$ (depending on the sign of the Fayet-Iliopoulos parameter $\theta$). We indeed find a single bosonic ground state in the correct representation of the global symmetry, meaning our result can be thought of as the first \emph{in silico} experimental confirmation of the mathematically predicted BPS spectrum of the model. Moreover, using the methods described in this note we have access to more than the count  of BPS states. This includes plots of the ground state and excited state wavefunctions, the gap with the first excited state as well as various field expectation values. The numerical methods we describe will hopefully make the study of the quantum nature of weakly coupled black hole bound states more tractable, and can potentially provide a useful experimental testing ground for other Witten index calculations.

The organization of this paper is as follows: in section \ref{setup}  we present the supercharges and symmetry generators of the theory. These fix the wavefunctions up to their dependence on radial variables. We also give a quick review of the stringy interpretation of this model. In section \ref{groundstatesec} we provide the BPS equations obeyed by the ground state wavefunction, suitably reduced via symmetries. We show that the (four-component) BPS wavefunction is fully determined by the solution to a single second-order partial differential equation in two variables, which we were unable to solve analytically. We also plot numerical solutions (BPS and non-BPS) for the full Schr\"{o}dinger problem in the singlet sector of the bosonic symmetry group. We study the dependence of the BPS and non-BPS wavefunctions on $\nu$. We conclude in section \ref{outlook}. We have collected formulae in appendices \ref{sphereapp} and \ref{wavefunceqs} and review the Born-Oppenheimer approximate ground state wavefunction in appendix \ref{boapp}.

\section{Setup}\label{setup}
\subsection{Supercharges}
In this paper we focus on the $\mathcal{N}=4$ supersymmetric quantum mechanics defined by the following four supercharges:
\begin{align}
Q_\alpha&\equiv\sqrt{2}\left[\mathds{1}\,\partial_{\bar{\phi}}-\phi\, \vec{\mathbf{x}}\cdot\vec{\boldsymbol{\sigma}}\right]_\alpha^{~\,\gamma}\left(\bar{\psi}\epsilon\right)_\gamma+i\left[\nabla_{\vec{\mathbf{x}}}\cdot\vec{\boldsymbol{\sigma}}-\left(|\phi|^2+\theta\right)\mathds{1}\right]_\alpha^{~\,\gamma}\lambda_\gamma~,\label{q}\\
\bar{Q}^\beta&\equiv-\sqrt{2}\left(\epsilon\psi\right)^\gamma\left[\mathds{1}\,\partial_{\phi}+\bar{\phi}\, \vec{\mathbf{x}}\cdot\vec{\boldsymbol{\sigma}}\right]_\gamma^{~\,\beta}+i\,\bar{\lambda}^\gamma\left[\nabla_{\vec{\mathbf{x}}}\cdot\vec{\boldsymbol{\sigma}}+\left(|\phi|^2+\theta\right)\mathds{1}\right]_\gamma^{~\,\beta}~.\label{qb}
\end{align}
We will discuss the stringy origin of this quantum mechanics in section \ref{stringy}, and refer the reader to \cite{Denef:2002ru} for a more in depth treatment. The Lagrangian of the model can also be found in \cite{Denef:2002ru}. In the above expressions, the $\boldsymbol{\sigma}^i$ are the usual Pauli matrices and $\vec{\boldsymbol{\sigma}}=\left\{\boldsymbol{\sigma}^1,\boldsymbol{\sigma}^2,\boldsymbol{\sigma}^3\right\}$. This quantum mechanics arises as the dimensional reduction of a $d=4$, $\mathcal{N}=1$, $U(1)$ gauge theory coupled to chiral matter\cite{Smilga:1986tv}. The matter content therefore includes a vector multiplet $(A,\,\vec{\mathbf{x}},\,\lambda_\alpha,\,\bar{\lambda}^\beta,D)$ and a chiral multiplet $(\phi,\,\psi_\alpha, F)$, along with its complex conjugate.\footnote{Notice that neither $D$ nor $F$ appear in the supercharges. This is because they are replaced by their on-shell values, e.g. $D_{\rm on-shell}=\frac{|\phi|^2+\theta}{\mu}$ and $F_{\rm on-shell}=\frac{1}{2}\frac{\partial W}{\partial \phi}=0$, when going to the operator formalism. The model under consideration has no superpotential and hence $W=0$.} The constant $\theta$ is a Fayet-Iliopoulos parameter. The triplet $\vec{\mathbf{x}}=\{\mathbf{x}^1,\mathbf{x}^2,\mathbf{x}^3\}$ transforms as a vector under $SO(3)$ rotations. The $d=1$ $U(1)$ gauge connection $A$ is non-dynamical and thus neither appears in the supercharges nor in the Hamiltonian. However, as is manifestly the case in the Lagrangian formulation of this theory, the chiral fields are charged under $A$, and therefore $A$ generates a Gauss-law constraint, which will appear in the algebra below.  

The two-component chiral fermion $\psi_\alpha$ and gaugino $\lambda_\alpha$ obey the canonical quantization conditions:
\begin{equation}\label{ccrm}
\left\{\psi_\alpha,\bar{\psi}^{\beta}\right\}=\delta_{\alpha}^{~\,\beta}~,\quad\quad\quad\quad\left\{\lambda_\alpha,\bar{\lambda}^{\beta}\right\}=\tfrac{1}{\mu}\delta_{\alpha}^{~\,\beta}~.
\end{equation}
We have borrowed the notation of \cite{Denef:2002ru} where spinors with an index down transform in the $\boldsymbol{2}$ of $SO(3)$ and spinors with an index up transform in the $\bar{\boldsymbol{2}}$. Indices are raised and lowered using the Levi-Civita symbol $\epsilon^{\alpha\beta}=-\epsilon_{\alpha\beta}$ with $\epsilon^{12}=1$. Thus in our conventions:
\begin{equation}
\left(\bar{\psi}\epsilon\right)_\alpha=\bar{\psi}^{\gamma}\epsilon_{\gamma\alpha}~,\quad\quad\quad\quad \left(\epsilon\psi\right)^\alpha=\epsilon^{\alpha\gamma}\psi_\gamma~,\quad\quad\quad\quad \epsilon_{\alpha\omega}\epsilon^{\omega\beta}=\delta_\alpha^{~\,\beta}~.
\end{equation}
Whenever spinor indices are suppressed it indicates that they are summed over.

The supercharges (\ref{q}-\ref{qb}) satisfy the supersymmetry algebra
\begin{equation}
\left\{Q_\alpha,\bar{Q}^\beta\right\}=2\left(\delta_\alpha^{~\,\beta} H -\vec{\mathbf{x}}\cdot\vec{\boldsymbol{\sigma}}_{\alpha}^{~\,\beta} G\right)~,\quad\quad\quad\quad\Big\{Q_\alpha,Q_\beta\Big\}=\Big\{\bar{Q}^\alpha,\bar{Q}^\beta\Big\}=0~,
\end{equation}
with Hamiltonian
\begin{equation}\label{ham}
H\equiv-\partial_{\phi}\partial_{\bar{\phi}}-\frac{1}{2\mu}\nabla_{\vec{\mathbf{x}}}^2+\frac{(|\phi|^2+\theta)^2}{2\mu}+\vec{\mathbf{x}}^2\,|\phi|^2+\bar{\psi}\,\vec{\mathbf{x}}\cdot\vec{\boldsymbol{\sigma}}\,\psi+\sqrt{2}\,i\left(\phi\,\left(\bar{\psi}\epsilon\right)\bar{\lambda}-\lambda\left(\epsilon\psi\right)\bar{\phi}\right)~.
\end{equation}
The operator
\begin{equation}
G\equiv\bar{\phi}\,\partial_{\bar{\phi}}-\phi\,\partial_{\phi}+\bar{\psi}\psi-1
\end{equation}
generates the Gauss-law constraint, and must vanish on gauge-invariant states.\footnote{Note that $G$ appears in the supersymmetry algebra with the normal ordering constant $-1$, which is not present in the classical generator.}  Thus, the proper supersymmetric states of the theory satisfy the following BPS equations
\begin{equation}\label{bpseqs}
 Q_\alpha\, \Psi=\bar{Q}^\alpha\, \Psi=0~, \quad\quad\quad\quad\quad G\,\Psi=0~.
\end{equation}

Let us briefly list the dimensions of the parameters and fields in units of the energy $[\mathcal{E}]$. They are  $[\phi]=-1/2$, $[\vec{\mathbf{x}}]=1$, $[\psi]=0$, $[\lambda]={3/2}$, $[\mu]=-3$ and $[\theta]=-1$, where the right hand sides of these expressions are shorthand for powers of $[\mathcal{E}]$. An important role will be played by the dimensionless quantity
\begin{equation}\label{nu}
\nu\equiv\frac{\mu^{1/3}}{|\theta|}~
\end{equation}
which we note here for later use.

\subsection{Conserved quantities}
The symmetry group of this theory is $SU(2)_J\times U(1)_R$ \cite{Cordova:2014oxa,Hori:2014tda} and the states fall into representations of this symmetry group. We will use this to our advantage. The generator of $SU(2)_J$ is the angular momentum operator:
\begin{equation}
\vec{\mathbf{J}}\equiv\vec{\mathbf{L}}+\frac{1}{2}\bar{\psi}\,\vec{\boldsymbol{\sigma}}\,\psi+\frac{\mu}{2}\bar{\lambda}\,\vec{\boldsymbol{\sigma}}\,\lambda~,
\end{equation}
with $\vec{\mathbf{L}}\equiv-i\,\vec{\mathbf{x}}\,\times\nabla_{\vec{\mathbf{x}}}$ the usual orbital angular momentum, which is not independently conserved. The components of $\vec{\mathbf{J}}$ satisfy the following algebra:
\begin{align}
&\left[\mathbf{J}^i,\mathbf{J}^j\right]=i\,\epsilon_{ijk}\,\mathbf{J}^k~, &&\left[\mathbf{J}^i,Q_\alpha\right]=-\frac{1}{2}\boldsymbol{\sigma}^{i~\gamma}_\alpha Q_\gamma~,\nonumber\\
&\left[\vec{\mathbf{J}}^2,\mathbf{J}^i\right]=0~, &&\left[\mathbf{J}^i,\bar{Q}^\alpha\right]=\frac{1}{2}\bar{Q}^\beta\,\boldsymbol{\sigma}^{i~\alpha}_\beta ~.\label{jalg}
\end{align}
Since $H=\tfrac{1}{4}\left\{Q_\alpha,\bar{Q}^\alpha\right\}$, it is easy to verify that (\ref{jalg}) implies $\left[\mathbf{J}^i,H\right]=0$ as promised.

The $U(1)_R$ generator \footnote{If we had instead considered $k$ identical chiral multiplets labeled by the index $a=1,\dots,k$ then the $R$-charge operator would take the form \begin{equation}
R\equiv \mu\,\bar{\lambda}\lambda+\sum_{a=1}^kq^a\left(\bar{\phi}^a\partial_{\bar{\phi}^a}-\phi^a\partial_{\phi^a}\right)+\left(q^a-1\right)\bar{\psi}^{a}\psi^a~,\nonumber
\end{equation}with the $\left\{q^a\right\}$ an arbitrary set of numbers. Since herein we are considering the case $k=1$, the term proportional to $q^1$ is simply $G+1$ and shifts the $R$-charge operator by the constant $q^1$ on gauge-invariant states, hence we set $q^1=0$. }
\begin{equation}\label{rcharge}
R\equiv \mu\,\bar{\lambda}\lambda-\bar{\psi}\psi~,
\end{equation}
satisfies
\begin{equation}
[R,Q_\alpha]=-Q_\alpha~,\quad\quad\quad\quad\left[R,\bar{Q}^\alpha\right]=+\bar{Q}^\alpha~,\quad\quad\quad\quad\left[R,\mathbf{J}^i\right]=0~.\label{ralg}
\end{equation}
Again we may deduce that $[R,H]=0$. Notice that the conserved $R$-charge counts the difference in the number of gauginos and chiral fermions, while the fermion number $F\equiv \mu\,\bar{\lambda}\lambda+\bar{\psi}\psi$ is not conserved as a result of the mixing term in (\ref{ham}).

The Hamiltonian is also invariant under the following discrete symmetry:
\begin{equation}\label{discretesymm}
\left(\bar{\psi}\epsilon\right)_\alpha\rightarrow \psi_\alpha~, \quad\quad \left(\epsilon\psi\right)^\alpha\rightarrow \bar{\psi}^\alpha~,\quad\quad
\bar{\lambda}^\alpha\rightarrow \left(\epsilon\lambda\right)^\alpha~,\quad\quad \lambda_\alpha\rightarrow \left(\bar{\lambda}\epsilon\right)_\alpha~,\quad\quad
\bar{\phi}\rightarrow -\phi~,\quad\quad \phi\rightarrow-\bar{\phi}~,
\end{equation}
which takes $R\rightarrow-R$. As a result the spectrum is degenerate under the interchange $R\leftrightarrow -R$, and we verify this explicitly in appendix \ref{wavefunceqs}.

\subsection{Stringy interpretation of the model and motivation}\label{stringy}
This simple quantum mechanical model can be understood as the dimensional reduction of the worldvolume gauge theory living on a pair of wrapped D-branes\footnote{D3-branes in type IIB string theory or D2p-branes in type IIA string theory.} in a Calabi-Yau compactification down to four dimensions. We consider here only the relative gauge group between the branes and ignore the ``center of mass'' degrees of freedom as they play no role in the study of the bound states. The vector muliplet scalars $\vec{\mathbf{x}}$ parametrize the separation between the D-branes in the non-compact space. Furthermore, in our particular setup, the two branes intersect at a single point in the Calabi-Yau, where there exists a stretched string between them \cite{Berkooz:1996km}. The lightest string excitation is parametrized by the bifundamental charged chiral multiplet and the chiral scalar mass can be read off from the classical potential
\begin{equation}\label{potential}
V\equiv\frac{1}{2\mu}\left(|\phi|^2+\theta\right)^2+\vec{\mathbf{x}}^2\,|\phi|^2~,
\end{equation}
that is $m_\phi^2=\vec{\mathbf{x}}^2+\theta/\mu$~. The quantity $\theta/\mu$ should be understood as the difference between the phases of the complexified masses of the D-branes in the language of the Calabi-Yau compactification. Thus,  having neglected the tower of higher string modes, this model is only valid as an approximation to the D-brane system when $|\vec{\mathbf{x}}|\ll l_s$ and $|\theta/\mu|^{1/2}\ll l_s$ where $l_s$ is the string length (see \cite{Denef:2002ru, Denef:2007vg} for an in-depth discussion as well as \cite{Berkooz:1996km,Kachru:1999vj} for earlier references).

An important role is played by the Fayet-Iliopoulos parameter $\theta$, which determines whether or not supersymmetry is broken, as is evidenced in (\ref{potential}). For $\theta<0$ there are two types of classical minima : the Higgs branch with $\left\{|\phi|^2=-\theta,~|\vec{\mathbf{x}}|=0\right\}$ (modulo $U(1)$ gauge transformations), which preserves supersymmetry; and the Coulomb branch with $\left\{|\phi|=0,~\vec{\mathbf{x}}^2 >-\theta/\mu\right\}$, where supersymmetry is classically broken. The potential has a flat direction when $|\phi|=0$ for which $V=\theta^2/2\mu$. Thus for $\theta<0$ the model is expected to have bound states with energies below this threshold value of $\theta^2/2\mu$. For $\theta>0$ there is no Higgs branch and the Coulomb branch is metastable for all values of $|\vec{\mathbf{x}}|$.

This picture changes drastically (for $\theta<0$) when quantum effects are taken into consideration. If we are interested in the large distance behavior of the wavefunctions of this model, that is $\vec{\mathbf{x}}^2\gg -\theta/\mu$, then the chiral multiplet can be integrated out (see appendix \ref{boapp}) generating an effective potential on the vector multiplet
\begin{equation}
V^{\rm eff}=\frac{1}{2\mu}\left(\frac{1}{2|\vec{\mathbf{x}}|}+\theta\right)^2~.
\end{equation}
The new effective potential has a zero energy minimum at $|\vec{\mathbf{x}}|=-1/2\theta$ and supersymmetry is restored on the Coulomb branch. For the Coulomb branch description to be consistent we thus need $\sqrt{\frac{|\theta|}{\mu}}\ll\frac{1}{|\theta|}$ or, restated in the language of the dimensionless parameter $\nu$ given in (\ref{nu}):
\begin{equation}
1\ll \nu~.
\end{equation}
For $\nu<1$ the BPS wavefunction is localized on the Higgs branch and the Coulomb branch description is no longer valid. For $\nu$ sufficiently large we expect the quiver description to break down and for supergravity to become the valid description \cite{Denef:2002ru}. Thus, in this model $\nu$ plays the role of the closed string coupling $g_s$. Since this model is a dimensionally reduced worldvolume gauge theory, the mass $\mu$ of the vector multiplet is related to the closed string coupling $\mu\propto 1/g_s$.

It is our interest in this paper to study how the quantum states of this quiver model depend on $\nu$. In quantum mechanics the Higgs and Coulomb branches are not fully separate, and the wavefunction will always have support in both directions. We aim to answer how the majority of the wavefunction's support moves from the $|\phi|$ direction to the $|\vec{\mathbf{x}}|$ direction as $\nu$ is increased. In order to study this we resort to numerically solving the full Schr\"{o}dinger equation and obtain supersymmetric and non-supersymmetric wavefunctions. This also serves as a numerical check of known supersymmetric index calculations \cite{Cordova:2014oxa,Hori:2014tda,Ohta:2014ria,Kim:2015oxa}. In doing so we also gain insight into the excited state spectrum of the theory and show how the gap with the first excited state depends on $\nu$.

\section{R=0 sector and the supersymmetric ground state}\label{groundstatesec}
We wish to diagonalize the maximal set of mutually commuting operators: $H,\, R,\, \mathbf{J}^3$ and $\vec{\mathbf{J}}^2$~. Before doing so, it is convenient to introduce the following parametrization for the chiral bosons:
\begin{equation}
\phi=\frac{\tilde{r}}{\sqrt{2}}e^{i\gamma}~,\quad\quad\quad\quad\quad \bar{\phi}=\frac{\tilde{r}}{\sqrt{2}}e^{-i\gamma}~.
\end{equation}
Derivatives with respect to $\phi$ and $\bar{\phi}$ are now given by
\begin{equation}
\partial_\phi=\frac{e^{-i\gamma}}{\sqrt{2}}\left(\partial_{\tilde{r}}-\frac{i}{\tilde{r}}\partial_\gamma\right)~,\quad\quad\quad\partial_{\bar{\phi}}=\frac{e^{i\gamma}}{\sqrt{2}}\left(\partial_{\tilde{r}}+\frac{i}{\tilde{r}}\partial_\gamma\right)~,\quad\quad\quad \partial_\phi\partial_{\bar{\phi}}=\frac{1}{2}\left(\frac{1}{\tilde{r}}\partial_{\tilde{r}}\,\tilde{r}\,\partial_{\tilde{r}}+\frac{1}{\tilde{r}^2}\partial_\gamma^2\right)~,
\end{equation}
and the Gauss law generator becomes
\begin{equation}
G=i\,\partial_\gamma+\bar{\psi}\psi-1~.
\end{equation}
 We will also find it convenient to work in spherical coordinates for the vector multiplet degrees of freedom, that is $\vec{\mathbf{x}}\rightarrow(r,\vartheta,\varphi)$. Details on how operators are represented in spherical coordinates are given in appendix \ref{sphereapp}.

\subsection{BPS equations}
Let us first attempt to solve the BPS equations (\ref{bpseqs}) for the ground state wavefunctions of this model. Usually this would involve applying the supercharges to a generic state and looking for the normalizable solutions. It is to our advantage, however, that the refined supersymmetric index \cite{Cordova:2014oxa,Hori:2014tda,Ohta:2014ria,Kim:2015oxa}
 \begin{equation}
 W_I\equiv\text{Tr}_{\mathcal H}\left\lbrace\, (-1)^{2\mathbf{J}^3}\,e^{-\beta\,H}y^{R+2\mathbf{J}^3}\right\rbrace
 \end{equation}
has been computed for this system and evaluates to $+1$ when $\theta<0$. As $\theta$ goes from negative to positive, $W_I$ jumps to zero and provides the simplest example of wall crossing \cite{Denef:2002ru,Denef:2007vg,Hori:2014tda}. Since $W_I$ is independent of $y$ there exists a single bosonic ground state with $R=0$ and $\mathbf{J}^3=0$.

The most general gauge-invariant, $R$-symmetric wavefunctions take the form:\footnote{We relegate the discussion on the wavefunctions in the other $R$-charge sectors to appendix \ref{wavefunceqs}.}
\begin{equation}\label{r0wf}
\Psi_0=\left\{e^{-i\gamma}D+E_{\alpha\beta}\,\bar{\psi}^\alpha\bar{\lambda}^\beta+e^{i\gamma}F\,\bar{\psi}^1\bar{\psi}^2\bar{\lambda}^1\bar{\lambda}^2\right\}|0\rangle~,
\end{equation}
where $D$, $E_{\alpha\beta}$ and $F$ are functions of $(\tilde{r},\vec{\mathbf{x}})$. We now further restrict to an $SU(2)_J$ highest weight state, which satisfies $\mathbf{J}^3\Psi_0=j\,\Psi_0$ and $\mathbf{J}^+\Psi_0=0$, yielding \footnote{We have introduced factors of $i$ in $D$ and $F$ as well as the factor of $\mu^{1/2}$ in $F$ for later convenience. With this choice the Schr\"{o}dinger equation is manifestly real.}
\begin{align}
D&=i\,e^{ij\varphi}\sin^j\vartheta\,\tilde{D}(\tilde{r},r)~,\quad\quad\quad F=i\,\mu^{1/2}\,e^{ij\varphi}\sin^j\vartheta\,\tilde{F}(\tilde{r},r)~,\\
E_{\alpha\beta}&=e^{ij\varphi}\sin^j\vartheta\,\begin{pmatrix}& e^{-i\varphi}\csc\vartheta\left[\tilde{E}_{11}+\cos\vartheta\left(\tilde{E}_{12}+\tilde{E}_{21}+\cos\vartheta\tilde{E}_{22}\right)\right] &\tilde{E}_{12}+\cos\vartheta\,\tilde{E}_{22}\\&\tilde{E}_{21}+\cos\vartheta\,\tilde{E}_{22} & e^{i\varphi}\sin\vartheta\,\tilde{E}_{22}\end{pmatrix}~,
\end{align}
where the $\tilde{E}_{ij}$ are also functions of $(\tilde{r},r)$. The ground state has spin $j=0$, meaning that $\Psi_0$ also satisfies $\mathbf{J}^-\Psi_0=0$, which leads to the further simplification
 \begin{equation}
 \tilde{E}_{21}=-\tilde{E}_{12}~,\quad\quad\quad\quad\quad\tilde{E}_{22}=-\tilde{E}_{11}~.
 \end{equation}
 Thus when $j=0$, the matrix $E_{\alpha\beta}$ is simply
 \begin{align}
E_{\alpha\beta}&=\begin{pmatrix}& e^{-i\varphi}\sin\vartheta\tilde{E}_{11} &\tilde{E}_{12}-\cos\vartheta\,\tilde{E}_{11}\\&-\tilde{E}_{12}-\cos\vartheta\,\tilde{E}_{11} & -e^{i\varphi}\sin\vartheta\,\tilde{E}_{11}\end{pmatrix}~.
\end{align}
Demanding $Q_\alpha\Psi_0=\bar{Q}^\beta\Psi_0=0$ leads to the following set of coupled differential equations
\begin{align}
 \tilde{D}-\frac{1}{
\mu^{1/2}}\tilde{F}&=0~,\label{deq}\\
\partial_{\tilde{r}}\,\tilde{E}_{12}+r\,\tilde{r}\,\tilde{E}_{11}-\frac{1}{\mu^{1/2}}\left(\frac{\tilde{r}^2}{2}+\theta\right)\tilde{F}&=0~,\label{e12eq1}\\
\partial_r\,\tilde{E}_{12} -r\,\tilde{r}\,\mu^{1/2}\,\tilde{F}-\left(\frac{\tilde{r}^2}{2}+\theta\right)\tilde{E}_{11}&=0~,\label{e12eq2}\\
\partial_{\tilde{r}}\,\tilde{E}_{11}+r\,\tilde{r}\tilde{E}_{12}-\frac{1}{\mu^{1/2}}\partial_r\,\tilde{F}&=0~,\label{e11eq1}\\
\frac{1}{r^2}\partial_r\left(r^2\,\tilde{E}_{11}\right)+\frac{\mu^{1/2}}{\tilde{r}}\partial_{\tilde{r}}\left(\tilde{r}\,\tilde{F}\right)-\left(\frac{\tilde{r}^2}{2}+\theta\right)\tilde{E}_{12}&=0~.\label{e11eq2}
\end{align}
We can solve for $\tilde{F}$ and $\tilde{E}_{11}$ algebraically in favor of $\tilde{E}_{12}$ using (\ref{e12eq1}) and (\ref{e12eq2}). Explicitly
\begin{align}
\tilde{F}&=\frac{1}{2\mu^{1/2}\,V}\left\lbrace r\,\tilde{r}\,\partial_r+\left(\frac{\tilde{r}^2}{2}+\theta\right)\partial_{\tilde{r}}\right\rbrace\tilde{E}_{12}~,\label{frel}\\
\tilde{E}_{11}&=-\frac{1}{2\,V}\left\lbrace r\,\tilde{r}\,\partial_{\tilde{r}}-\frac{1}{\mu}\left(\frac{\tilde{r}^2}{2}+\theta\right)\partial_r\right\rbrace \tilde{E}_{12}~,\label{e11rel}
\end{align}
where $V=\frac{\left(\frac{\tilde{r}^2}{2}+\theta\right)^2+\mu\,r^2\,\tilde{r}^2}{2\mu}$ is the classical potential, as in (\ref{potential}). Making these substitutions, we find that $\tilde{E}_{12}$ satisfies the following Schr\"{o}dinger equation:\footnote{The system (\ref{e12eq1}-\ref{e11eq2}) is overdetermined but not inconsistent. Substitution of (\ref{frel}) and (\ref{e11rel}) into both (\ref{e11eq1}) and (\ref{e11eq2}) gives rise to the same equation (\ref{hardschro}). }
\begin{equation}\label{hardschro}
-\frac{1}{2\mu\,r^2}\partial_r\left(r^2\,\partial_r\,\tilde{E}_{12}\right)-\frac{1}{2\,\tilde{r}}\partial_{\tilde{r}}\left(\tilde{r}\,\partial_{\tilde{r}}\,\tilde{E}_{12}\right)+\frac{2\,\tilde{r}}{\mu^{1/2}}\tilde{F}-r\tilde{E}_{11}+V\,\tilde{E}_{12}=0~.
\end{equation}
As we will see in (\ref{j0ham}), this is precisely what we get out of the full Schr\"{o}dinger equation, but here the BPS equations have provided us with expressions for $\tilde{F}$ and $\tilde{E}_{11}$ in terms of $\tilde{E}_{12}$, and therefore a single uncoupled partial differential equation.
Were we able to solve (\ref{hardschro}), we would have an analytic expression for the BPS wavefunction. This has proven too difficult, so we instead solve for the ground state numerically.  Before moving on we note that (\ref{hardschro}) is very much akin to the `deprolongation' of the BPS equations noticed in \cite{Sethi:2000ba}, namely we can rewrite (\ref{hardschro}) as
\begin{equation}
\left(-\mathcal{D}^2 +\vec{\mathbf{B}}\cdot\vec{\mathcal{D}}+V\right)\tilde{E}_{12}=0
\end{equation}
where we have implicitly defined the differential operator
\begin{equation}\label{laplaceop}
\mathcal{D}^2\equiv \frac{1}{2\mu\,r^2}\,\partial_r\,r^2\,\partial_r+\frac{1}{2\,\tilde{r}}\,\partial_{\tilde{r}}\,\tilde{r}\,\partial_{\tilde{r}}~,
\end{equation}
and
\begin{equation}
\vec{\mathbf{B}}\cdot\vec{\mathcal{D}}=\frac{1}{2\mu\,V}\left\lbrace2\,\tilde{r}\left(\frac{\tilde{r}^2+r^2\,\mu}{2}+\theta\right)\partial_{\tilde{r}}+r\left(\frac{3\,\tilde{r}^2}{2}-\theta\right)\partial_r\right\rbrace~.
\end{equation}

\subsection{Numerical setup and results}
To construct the ground state wavefunction numerically, we will input the full Schr\"{o}dinger problem into the \emph{Mathematica}'s {\tt NDEigensystem} command. The coupled differential equations of interest (see appendix \ref{r0eqsapp}) obtained from $H\Psi_0=\mathcal{E}\,\Psi_0$ are :
\begin{equation}\label{j0ham}
\begin{pmatrix}
-\mathcal{D}^2+V+\frac{1}{2\,\tilde{r}^2} &0 &0 &0 \\0& -\mathcal{D}^2+V+\frac{1}{2\tilde{r}^2} &0 & \frac{2\tilde{r}}{\mu^{1/2}}\\
0& 0 &-\mathcal{D}^2+V+\frac{1}{\mu\,r^2} &-r \\
\frac{\tilde{r}}{\mu^{1/2}} &\frac{2\tilde{r}}{\mu^{1/2}} &-r &-\mathcal{D}^2+V
\end{pmatrix}\begin{pmatrix}\Phi\\ \tilde{F}\\ \tilde{E}_{11}\\ \tilde{E}_{12}\end{pmatrix}=\mathcal{E}\begin{pmatrix}\Phi \\\tilde{F}\\ \tilde{E}_{11}\\ \tilde{E}_{12}\end{pmatrix}~,
\end{equation}
with $\mathcal{D}^2$ as in (\ref{laplaceop}). We have defined $\Phi\equiv\mu^{1/2}\,\tilde{D}-\tilde{F}$. By virtue of the BPS equations, $\Phi=0$ in the ground state, but it does not necessarily vanish in the excited states. It turns out that $\Phi$ satisfies the same wave equation as an $R=\pm 2$ state (see for example (\ref{hrm2}) and (\ref{hrp2})), hence states with vanishing $\Phi$ have no superpartners in the $R=\pm 2$ sectors. The excited states with $\Phi=0$ thus live in short multiplets, as described in appendix \ref{r0eqsapp}. We also note here that the norm of the $R=0,~j=0$ wavefunctions is given by
\begin{equation}
\langle \Psi_0^*,\Psi_0\rangle= \frac{1}{\mu}\int_0^\infty dr\int_0^{\infty}d\tilde{r}\int_0^\pi d\vartheta\int_0^{2\pi}d\varphi\int_0^{2\pi}d\gamma\,\tilde{r}\,r^2\,\sin\vartheta\left((\tilde{F}+\Phi)^2+\tilde{F}^2+2\left(\tilde{E}_{11}^2+\tilde{E}_{12}^2\right)\right)~.
\end{equation}
However, we will have to restrict the domain of $r$ and $\tilde{r}$ when computing the eigenfunctions of (\ref{j0ham}) numerically, and the wavefunctions will be normalized on this restricted domain. We are only interested in the case $\theta<0$, since this is the phase with unbroken supersymmetry and relegate the study of $\theta$ positive to later work.

The procedure is as follows, we input the differential operator (\ref{j0ham}) into \emph{Mathematica}'s {\tt NDEigensystem} command, which uses a finite element approach to solve for the eigenfunctions of a coupled differential operator on a restricted domain. We vary the size of the domain and the refinement of the mesh until the numerics are stable.

Before moving on we pause here to discuss the subtlety of diagnosing supersymmetry breaking in the presence of numerical errors. That is, it is naively quite difficult to tell the difference between an \emph{actual} zero energy eigenstate and one whose energy is lifted by supersymmetry breaking effects, especially when the algorithm's resolution of the energy is $\sim 10^{-3}-10^{-4}$ (as it can be in certain parameter regimes). This is where being able to numerically determine the excited state spectrum comes in handy. To diagnose if supersymmetry is broken, it suffices to check that the state of interest lives in a supersymmetric multiplet and to find its SUSY partners. If the state in question is a singlet under supersymmetry, then supersymmetry is unbroken.
\subsubsection{ $\nu<1$ and the Higgs branch}
\begin{figure}[t!]
 \centering
  \includegraphics[width=0.32\textwidth]{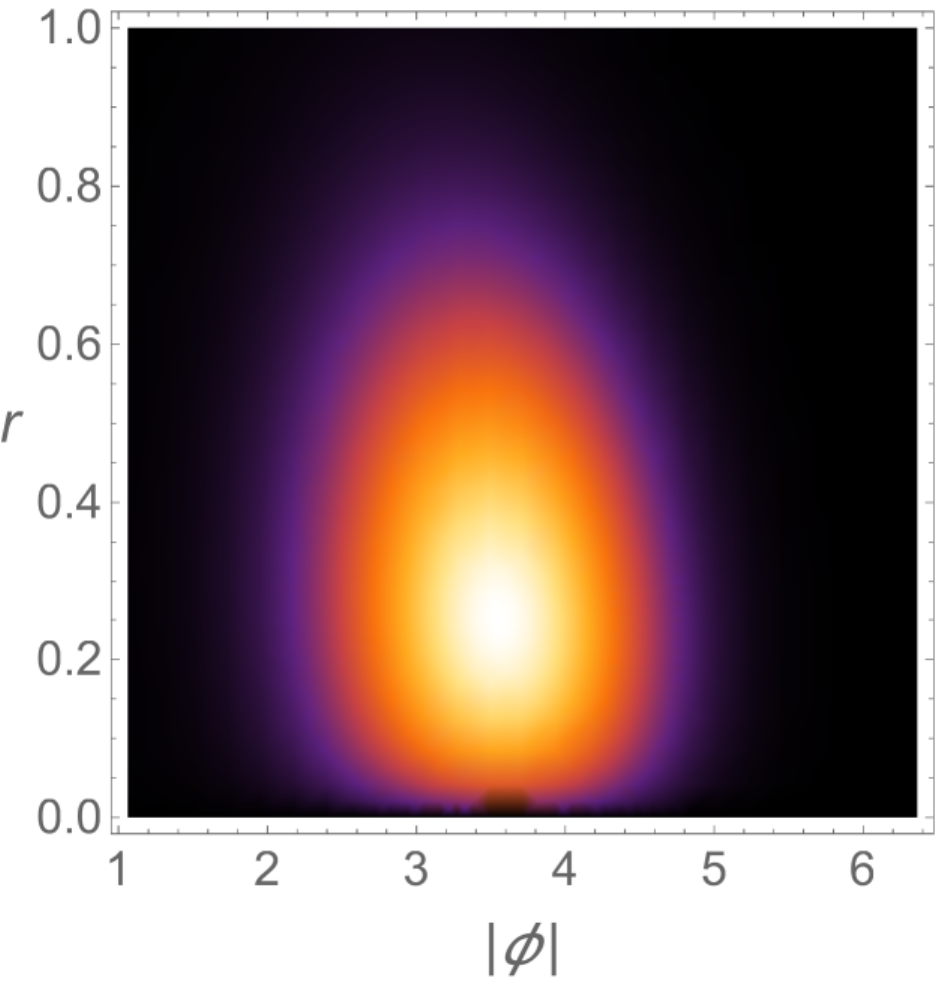}
  \includegraphics[width=0.32\textwidth]{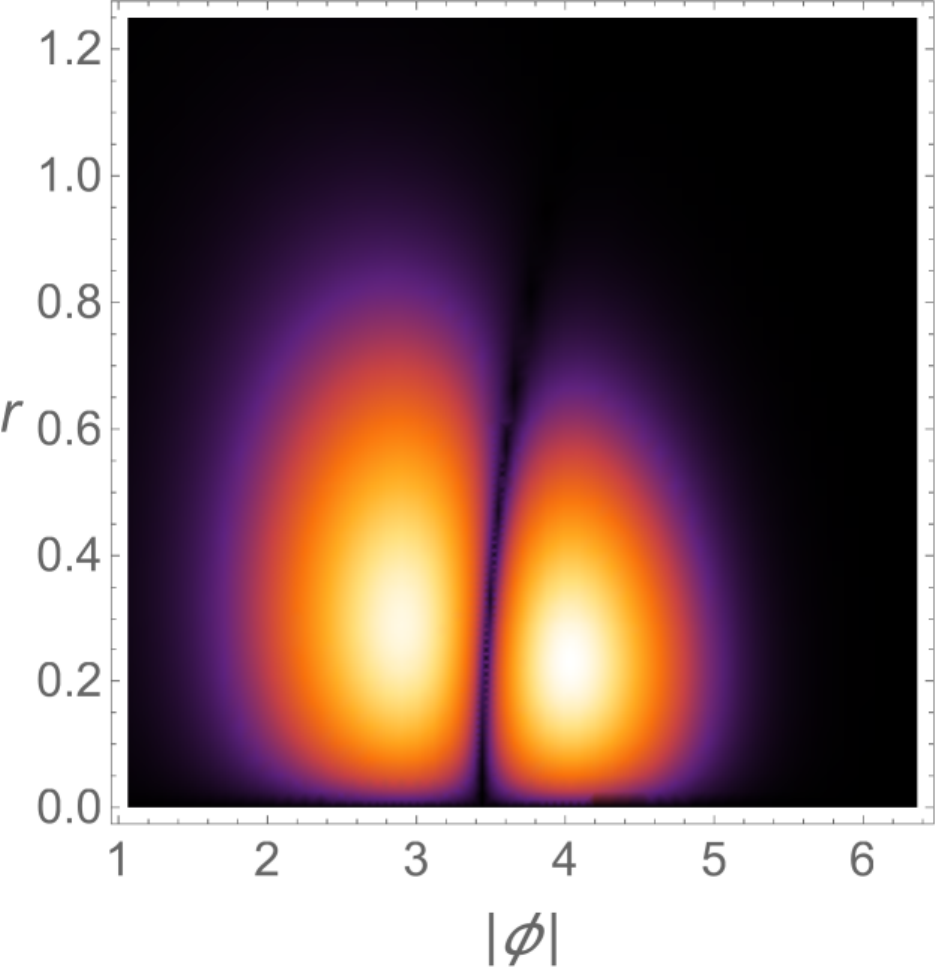}
  \includegraphics[width=0.32\textwidth]{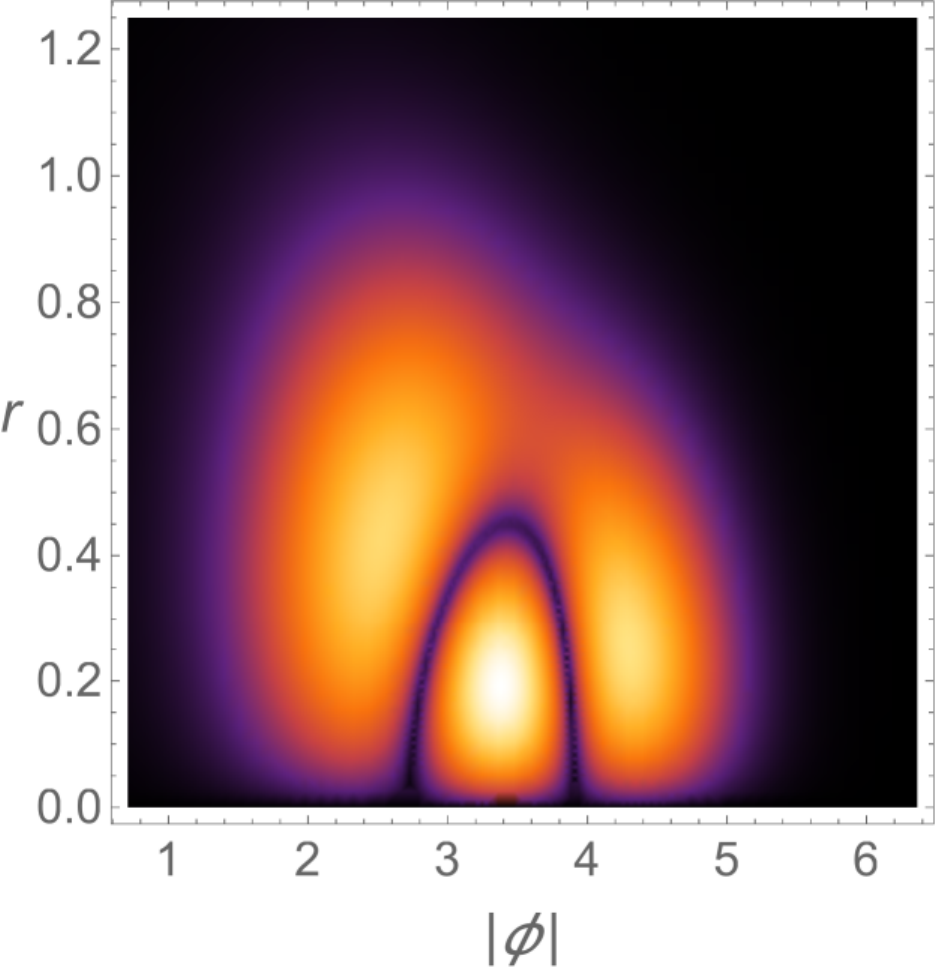}
   \includegraphics[width=0.32\textwidth]{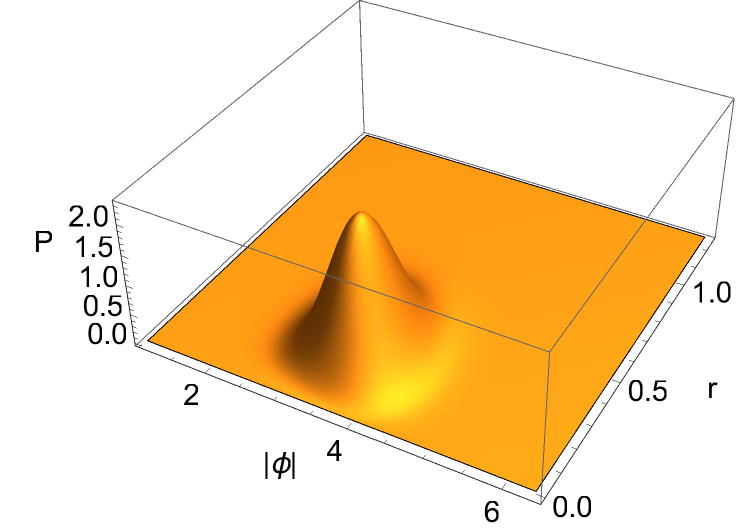}
  \includegraphics[width=0.32\textwidth]{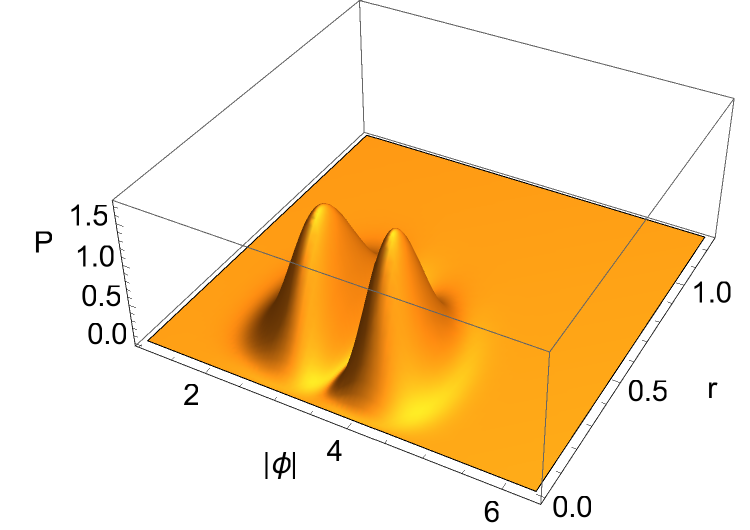}
  \includegraphics[width=0.32\textwidth]{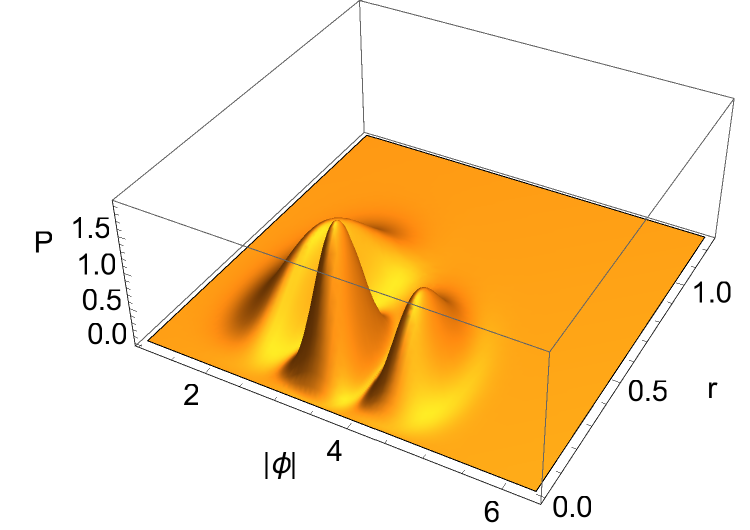}
\includegraphics[width=0.32\textwidth]{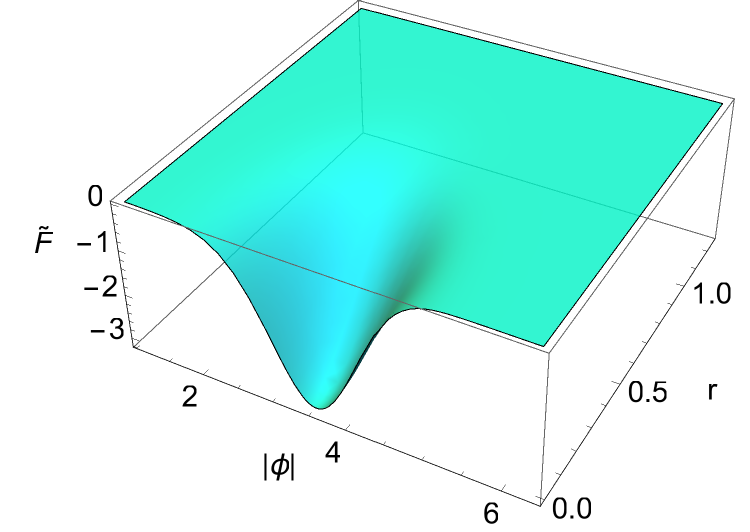}
  \includegraphics[width=0.32\textwidth]{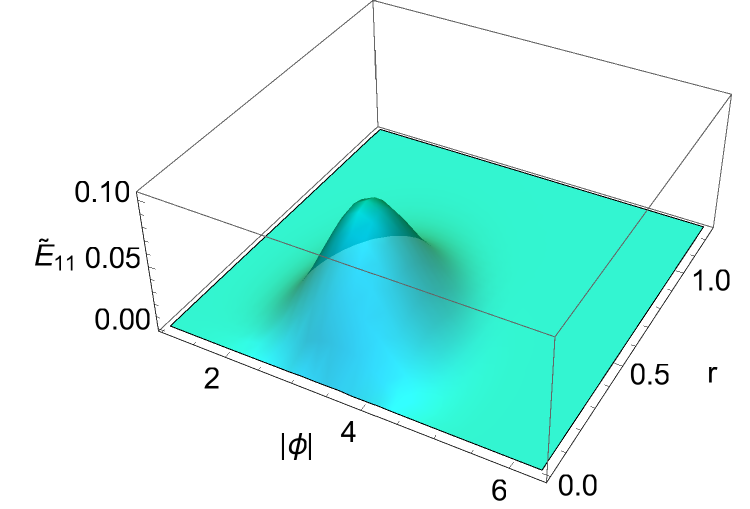}
  \includegraphics[width=0.32\textwidth]{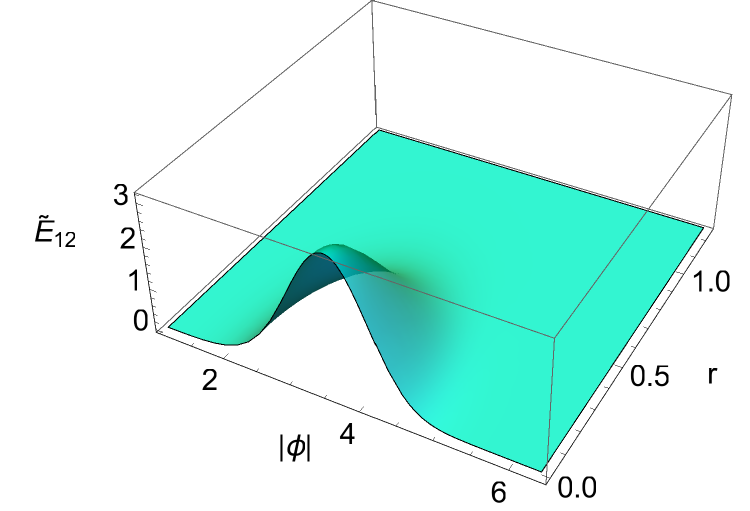}
  \caption{\emph{Top two rows}: Probability density for the first three energy eigenstates $P\sim \tilde{r}\,r^2|\Psi_0|^2$ (after integrating over the angular variables) in the $R=0$, $j=0$ sector with $\theta=-12.5,$ and $\mu=10$, corresponding to $\nu\approx 0.172$. The minimum of the potential is at $|\phi|\approx 3.53$.  The energies of the first three eigenstates are respectively $\mathcal{E}\approx (0,\,1.538,\,2.984)$. \emph{Bottom row}: Components of the ground state wavefunction.}\label{wfhiggs}
 \end{figure}

The radial probability density of the first three eigenstates in the $R=0,\,j=0$ sector are shown in figure \ref{wfhiggs} for a set of parameters such that $\nu\approx0.172$~. We also plot the components $(\tilde{F},\,\tilde{E}_{11},\,\tilde{E}_{12})$ of the ground state wavefunction individually. Let us summarize some results:

\begin{enumerate}
\item The ground state wavefunction passes several consistency checks. Note, for example, that it is peaked near the classical minimum $(|\phi|,r)=(\sqrt{-\theta},0)$. Furthermore, the ground state is nodeless, unlike the excited states, and has energy $\mathcal{E}\approx0$ within numerical error (that is, for a quick enough run we find $\mathcal{E}_0\sim 10^{-6}$).

\item To verify that we have actually found the supersymmetric ground state, it suffices to check that it is not part of a supersymmetric multiplet. That means we should check that it is not paired with any states in the $R=\pm1,\,j=1/2$ sectors. This is indeed the case. Furthermore, we were able to numerically find all the supersymmetric partners of the first excited state for the parameters shown in figure \ref{wfhiggs}. The multiplet consists of 8 states, 4 bosonic and 4 fermionic, which split into a  bosonic $R=0,\,j=0$ singlet, two fermionic doublets with $R=\pm 1,\, j=1/2$, and a bosonic triplet with $R=0,\,j=1$. In the $R=0,\,j=0$ sector, the first two excited state indeed satisfy $\Phi=0$.    
\item Generically in the ground state we find $\langle |\phi|^2\rangle\approx -\theta$, to a very good approximation (see figure \ref{expectationvals}). This is true independent of whether $\nu$ is less than or greater than 1. 
\item We fit the gap between the ground state and the first excited state and find that it is $\Delta\mathcal{E}\approx \sqrt{\frac{-2\theta}{\mu}}$, for $\nu\ll1$ (see figure \ref{expectationvals}).
\item Notice that the components of the wavefunction are all on equal footing. That is, they are all more or less peaked at the classical minimum, save for $\tilde{E}_{11}$ which vanishes at $r=0$ and is peaked slightly off the axis. It is worth mentioning however that their qualitative features are similar. We will find that this changes significantly on the Coulomb branch.
\end{enumerate}

\subsubsection{$\nu>1$ and the Coulomb branch}
The radial probability density for the first three eigenfunctions in the $R=0,\,j=0$ sector are shown in figure \ref{wfcoulomb} for a set of parameters such that $\nu\approx1.876$~. We again summarize some results:
\begin{figure}[t!]
 \centering
  \includegraphics[width=0.32\textwidth]{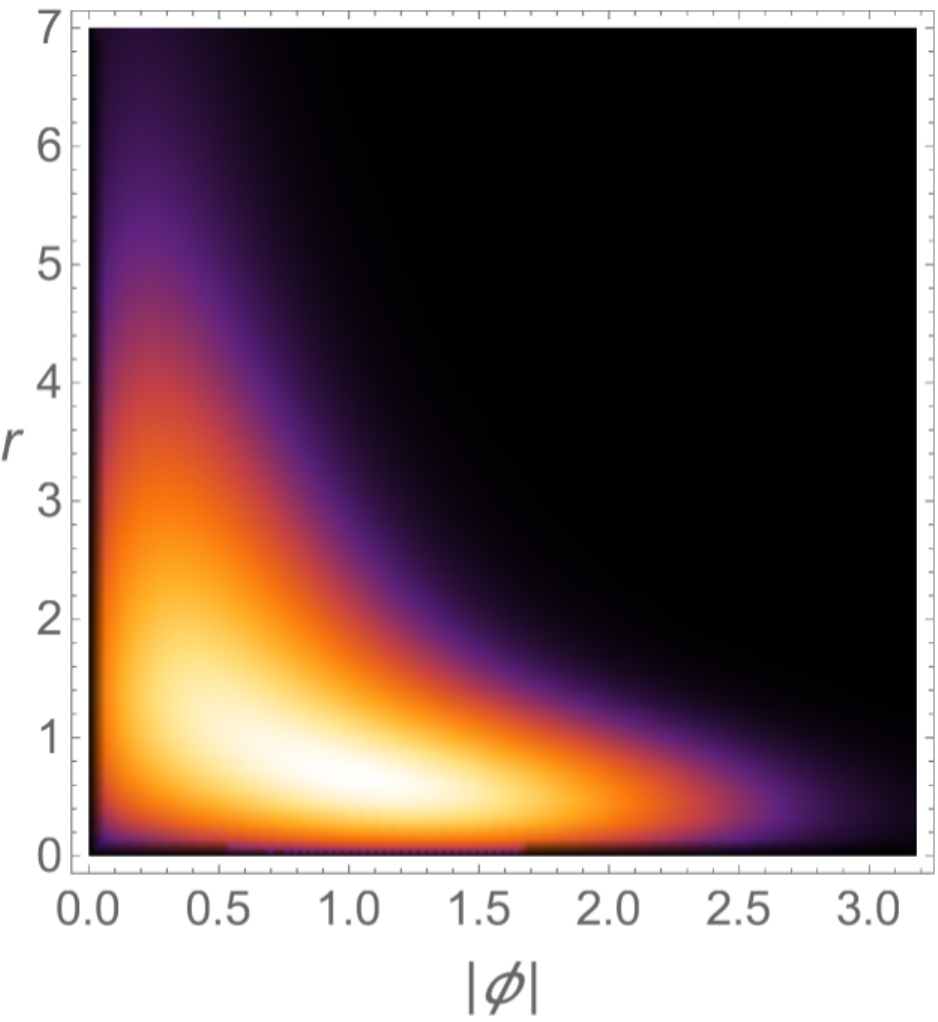}
  \includegraphics[width=0.32\textwidth]{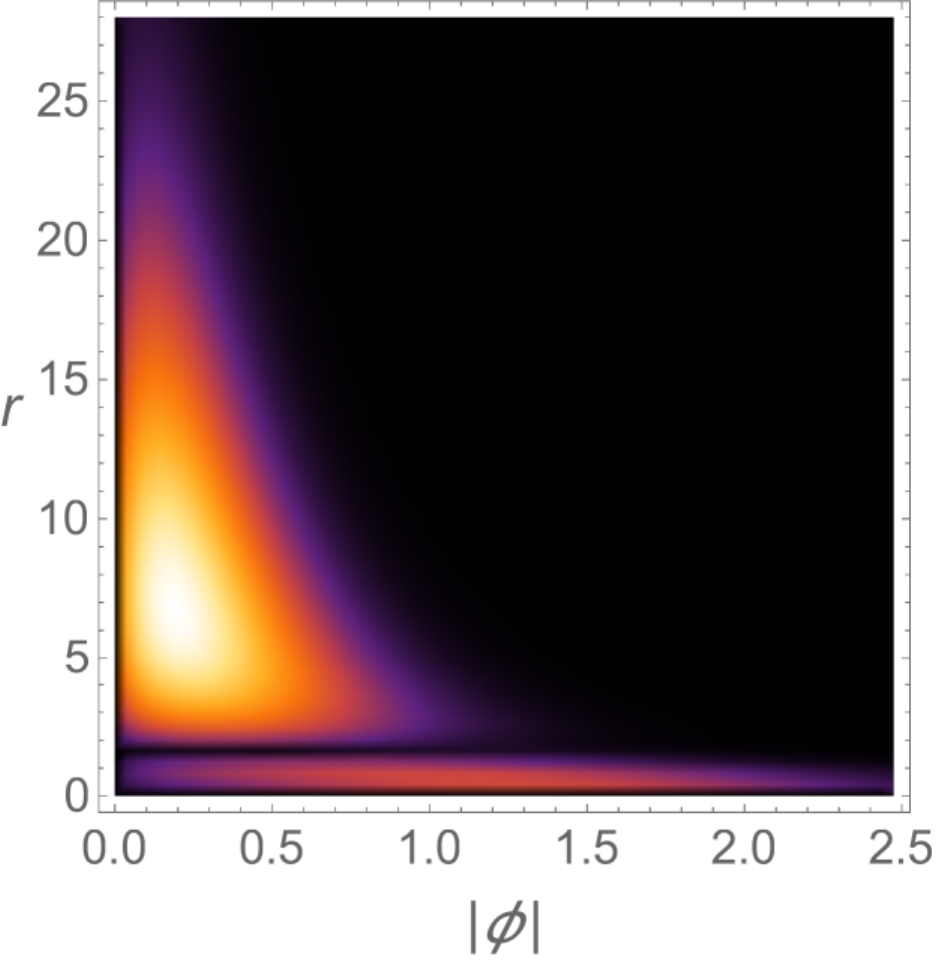}
  \includegraphics[width=0.32\textwidth]{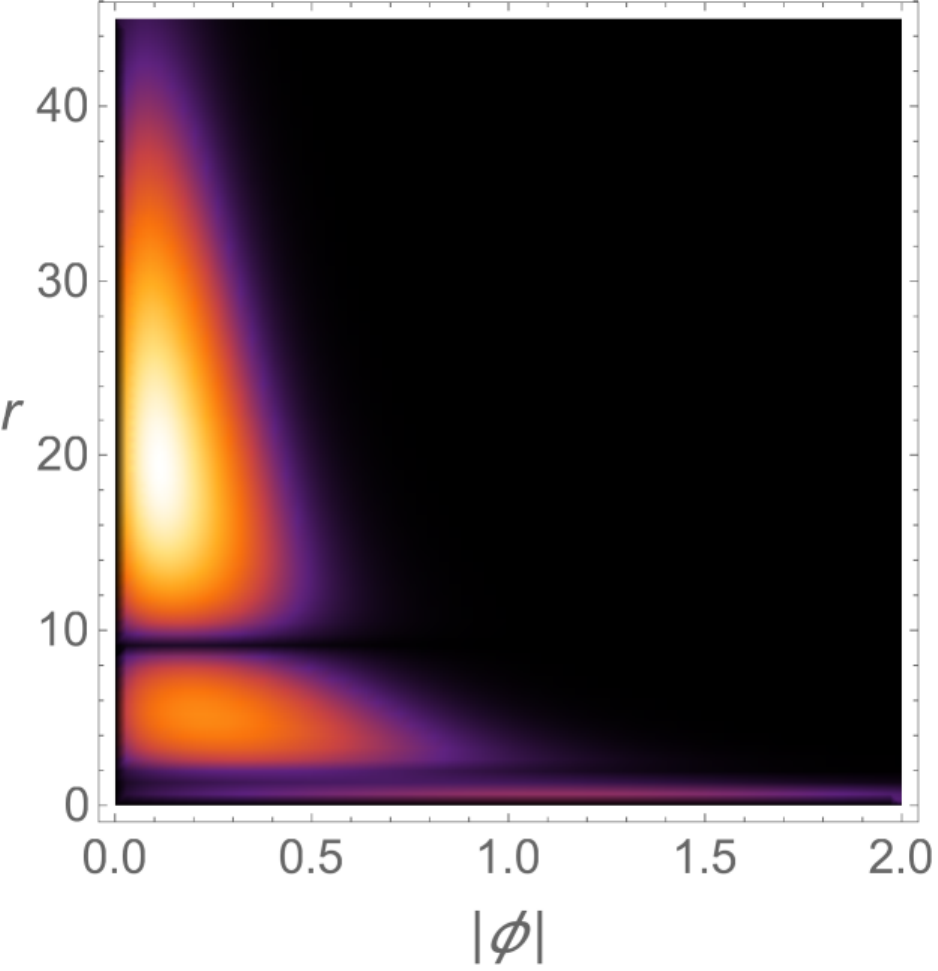}
   \includegraphics[width=0.32\textwidth]{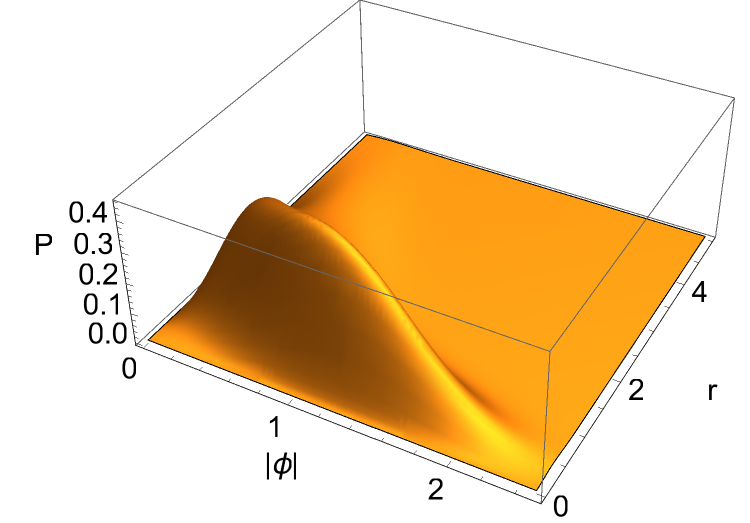}
  \includegraphics[width=0.32\textwidth]{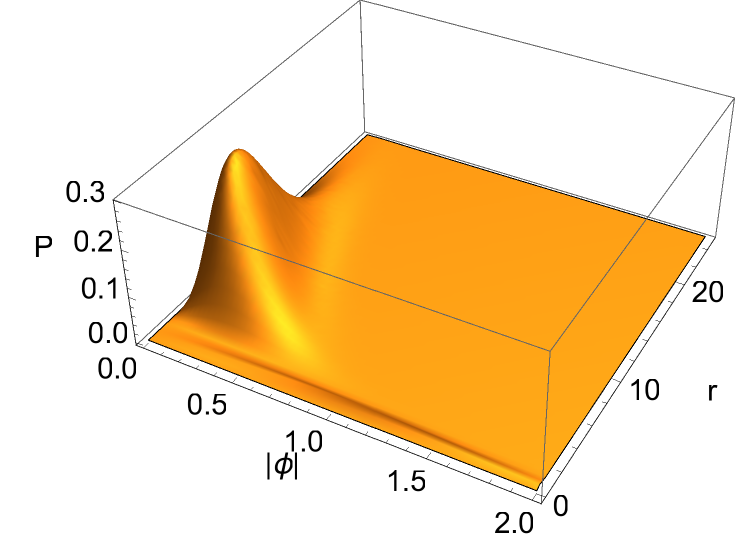}
  \includegraphics[width=0.32\textwidth]{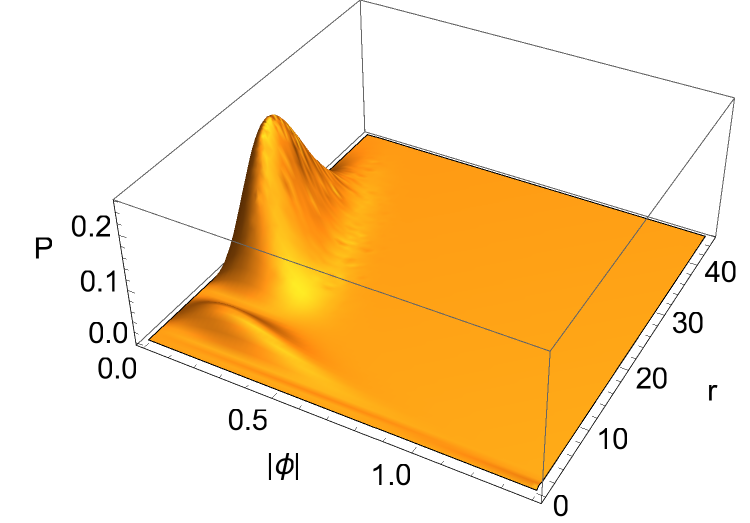}
\includegraphics[width=0.32\textwidth]{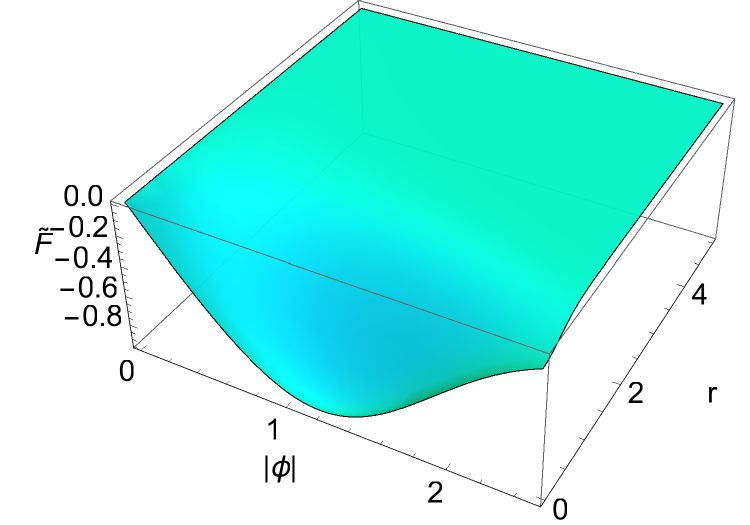}
  \includegraphics[width=0.32\textwidth]{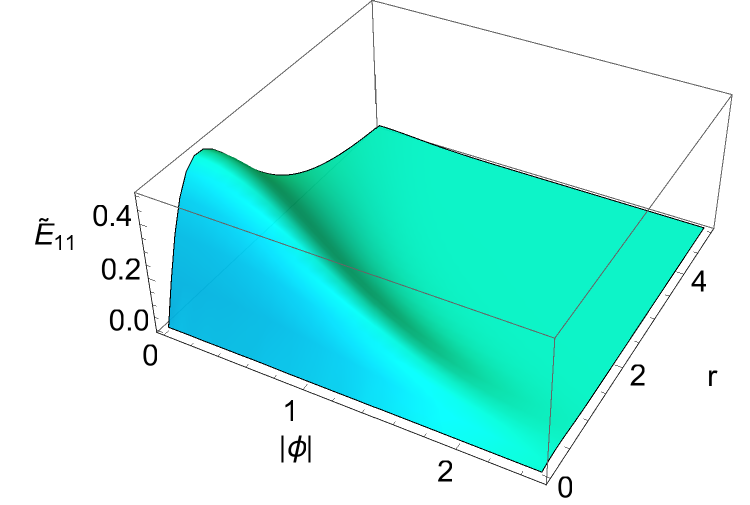}
  \includegraphics[width=0.32\textwidth]{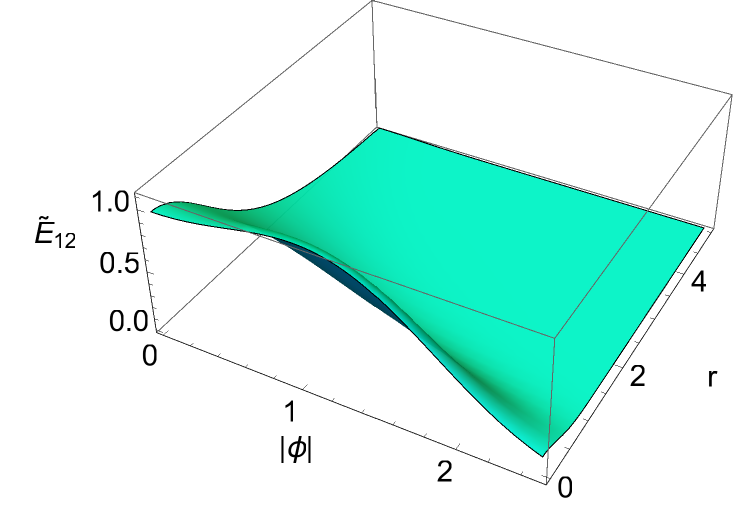}
  \caption
   {\emph{Top two rows}: Probability density for the first three energy eigenstates $P\sim \tilde{r}\,r^2|\Psi_0|^2$ (after integrating over the angular variables) in the $R=0$, $j=0$ sector for $\theta=-0.911$, $\mu=5$ corresponding to $\nu\approx 1.876$. The minimum of the potential is at $|\phi|\approx 0.955$. The minimum of $V^{\rm eff}$ is at $r\approx 0.55$ and the chiral fields become tachyonic for $r<r_h\equiv\sqrt{-\theta/\mu}\approx 0.427$. Note the difference in scales between the $r$-axis and the $|\phi|$-axis, particularly for the excited states.  The energies of the first three eigenstates are respectively $\mathcal{E}\approx(0,\,0.073,\,0.079)$. \emph{Bottom row}: Components of the ground state wavefunction.}\label{wfcoulomb}
\end{figure}
\begin{enumerate}
\item The main difference between the case at hand where $\nu>1$ and the previous case where $\nu<1$ is that the wavefunction has considerable support on the Coulomb branch, that is near $|\phi|=0$ and $r>r_h\equiv\sqrt{-\theta/\mu}$. Recall that the mass of $\phi$ is given by $m_\phi^2=r^2+\theta/\mu$, so $r_h$ is precisely where $\phi$ becomes tachyonic. The wavefunction's support on the Coulomb branch grows as $\nu$ gets larger (see figure \ref{probplot}). Notice how the ground state width in the $|\phi|$ direction increases as $r$ decreases. This can be understood from the Born-Oppenheimer analysis, where the wavefunction has a factor $\sim e^{-r|\phi|^2}$. For $r<r_h$ the Higgs branch becomes important and the Born-Oppenheimer analysis breaks down. We see that the width of the wavefunction in the $|\phi|$ direction grows greatly at this point.
\item The excited state wavefunctions have considerably less support on the Higgs branch than the ground state, and significantly more support on the Coulomb branch than the ground state.
\item The component $\tilde{F}$ of the wavefunction only has support near $r=0$. This is expected from the Born-Oppenheimer analysis given in appendix \ref{boapp}. Both $F$ and $D$ in (\ref{r0wf}) can be understood as excited states of the chiral Hamiltonian $H^{(0)}$ in (\ref{hamdecomp1}).
\item The exact energy spectrum on the Coulomb branch was calculated in \cite{Avery:2007xf} and is given by $\mathcal{E}_n=2n\frac{(n+1)}{(2n+1)^2}\frac{\theta^2}{\mu}$ for $n=0,1,\dots$. We confirm this by noting that the gap between the ground state and the first excited state in our numerically obtained spectrum fits well with $\Delta\mathcal{E}\approx\frac{4\,\theta^2}{9\mu}$ for $\nu$ sufficiently large (see figure \ref{expectationvals}). As $\nu$ increases the gap with the first excited state approaches zero.
\item In keeping with the Born-Oppenheimer analysis, we find that to a good approximation $\langle r\rangle \approx -1/\theta$ for $\nu$ sufficiently large. Again $\left\langle|\phi|^2\right\rangle=-\theta$.
\end{enumerate}

\subsubsection{Relating the two pictures}\label{relating}
\begin{figure}[t]
 \centering
  \includegraphics[width=0.48\textwidth]{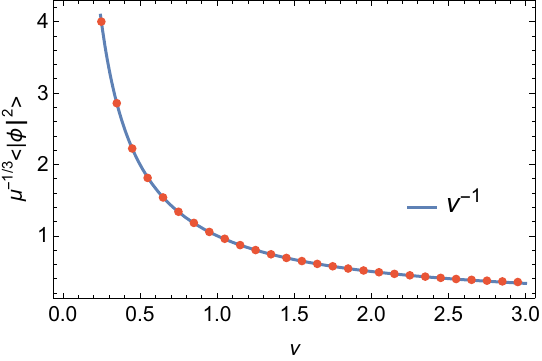}
  \includegraphics[width=0.48\textwidth]{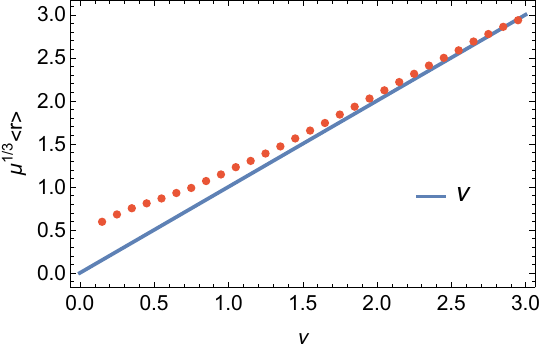}
\includegraphics[width=0.48\textwidth]{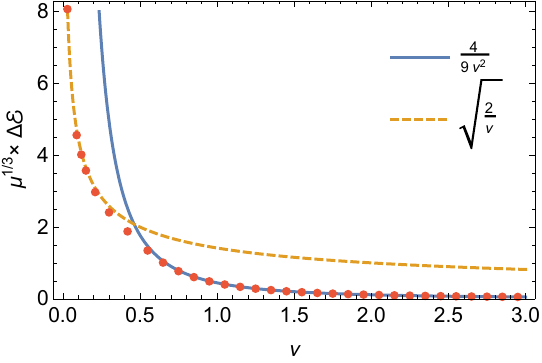}
  \includegraphics[width=0.48\textwidth]{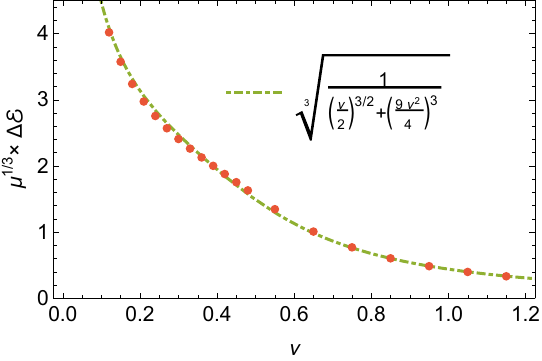}
  \caption {\emph{Top left}: Expectation value of $\mu^{-1/3}\left\langle|\phi|^2\right\rangle$ as a function of $\nu$. We conclude that $\left\langle|\phi|^2\right\rangle=-\theta$ in both the Higgs and Coulomb branch. \emph{Top right}: Expectation value of $\mu^{1/3}\left\langle r\right\rangle$ as a function of $\nu$. As $\nu$ increases the approximate Coulomb branch formula $\langle r\rangle=-1/\theta$ becomes more exact. \emph{Bottom row}: Gap between the ground state and the first excited state as a function of $\nu$. As $\nu$ increases we go from a regime where the gap is given by $\Delta\mathcal{E}\approx \sqrt{\frac{-2\theta}{\mu}}$ to a regime where the gap is $\Delta\mathcal{E}\approx \frac{4\,\theta^2}{9\mu}$.}\label{expectationvals}
\end{figure}

\begin{figure}[t]
 \centering
  \includegraphics[width=0.52\textwidth]{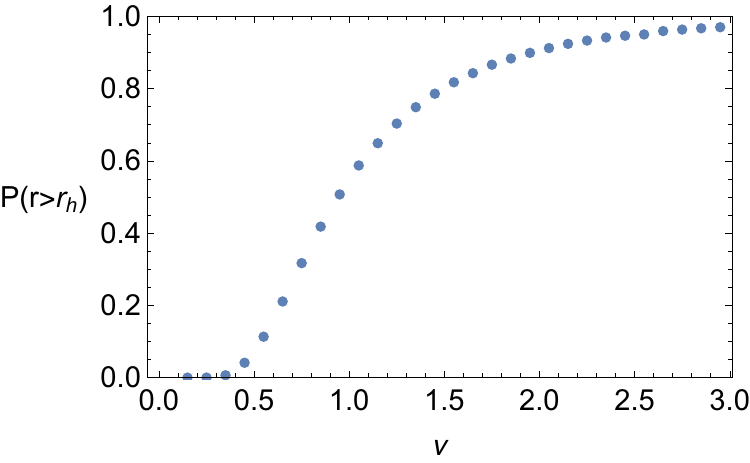}
  \caption {Fraction of the probability density, in the ground state, which has support for $r>r_h\equiv\sqrt{-\theta/\mu}$. }\label{probplot}
\end{figure}

We can now see that the ground state wavefunction morphs smoothly from having the bulk of its support on the Higgs branch to being supported mostly on the Coulomb branch as we increase $\nu$. This is accompanied by an increasing vev for $\langle r\rangle$ and a decreasing vev for $\left\langle|\phi^2|\right\rangle$. Since the ground state is nodeless, it must be the case that as $\nu$ is tuned larger ($|\theta|$ tuned smaller), the peak of the wavefunction moves towards the origin and the tails spread out in the L-shape seen in figure \ref{wfcoulomb}. This smooth metamorphosis is accompanied by the bulk of the probability being localized at large values of $r$. To make this quantitative, we compute the integrated probability
\begin{equation}
P(r>r_h)=\frac{\int_{r_h}^{\infty}dr\int_0^{\infty }d\tilde{r}\,\tilde{r}\,r^2 |\Psi_0|^2}{\int_{0}^{\infty}dr\int_0^{\infty}d\tilde{r}\,\tilde{r}\,r^2 |\Psi_0|^2~},
\end{equation}
where $r_h\equiv\sqrt{-\theta/\mu}$ is the naive value of $r$ where the Higgs branch becomes important, i.e. when the $\phi$ field becomes tachyonic. The function $P(r>r_h)$ is a diagnostic which tells us what fraction of the ground state wavefunction has support on the Coulomb branch. We have plotted it in figure \ref{probplot}.
As expected $P(r>r_h)$ increases monotonically as a function of $\nu$. For $\nu$ sufficiently large, the Born-Oppenheimer analysis, and the wavefunction given in appendix \ref{boapp} capture the bulk of the large $r$ physics. In particular the energy gap and the expectation of $\langle r\rangle$ match quite precisely with the Coulomb branch formulae for $\nu\gg1$, as shown in figure \ref{expectationvals}. It is also of note that tuning the Fayet-Illiopoulos parameter near the wall of marginal stability at $\theta\rightarrow 0^-$ forces us into the large $\nu$ regime. Hence to study wall crossing, one need only focus on the Coulomb branch \cite{Denef:2007vg}. Of course, the Higgs branch never fully decouples and the wavefunction will always have support there, it simply becomes less relevant as $\nu$ gets large as evidenced by the decreasing vev of $\left\langle|\phi|^2\right\rangle$.

For $\nu\sim 1$, there is no such decoupling, and we need to consider all degrees of freedom of the model. This case can be very rich. In more complicated quiver models for example, where there can be multiple vector multiplets and a large number of chiral multiplets in play, one can consider tuning parameters such that some subset of chiral multiplets and vector multiplets are relevant. It was advocated in \cite{Anninos:2013nra} that such a mixed Higgs-Coulomb problem would be a toy model for studying a D-particle falling into a black hole. In that case, where we have little analytic control, numerics seems to be one avenue toward approaching this problem. This seems exciting, as it would provide a numerical avenue towards testing the fuzzball proposal as advocated in \cite{Bena:2012zi}.

\section{Outlook}\label{outlook}

In this brief note we numerically constructed the bound state wavefunctions of a simple D-brane quiver quantum mechanics. In doing so we demonstrated how the ground state and excited states shuffle around as parameters are tuned. We have also given a numerical-experimental verification of the Witten index computed for this model \cite{Cordova:2014oxa,Hori:2014tda,Ohta:2014ria,Kim:2015oxa}. This is an example of what may someday be commonplace---an \emph{in silico} approach to studying supersymmetric quantum systems. Apart from numerically verifying complicated Witten index computations, one can also imagine, for example, revisiting old problems related to D0-brane matrix quantum mechanics \cite{Yi:1997eg,Moore:1998et,Plefka:1997xq,Porrati:1997ej,Frohlich:1999zf,Lin:2014wka} such as the structure of the D0-brane bound state at small distance for fixed $N$ or a non-Abelian quiver quantum mechanics at fixed but nonzero $N$. These systems have threshold bound states in the absence of Fayet-Iliopoulos parameters and we hope to report on whether numerics are useful in this scenario in a subsequent publication.  Another avenue is to study the states of mass deformed quiver theories recently described in \cite{Asplund:2015yda}, which represent the effective dynamics of wrapped branes in AdS. These questions are within reach using numerics.

Another interesting future direction is to study the entanglement structure between the chiral and vector degrees of freedom. Typically when studying these models we have recourse to the Born-Oppenheimer approximation in which the chiral degrees of freedom are integrated out. As we mentioned, this integrating out generates a metric on the moduli space of the Coulomb branch and it would be interesting to understand if this metric can be linked to the entanglement between the degrees of freedom.

Left untouched of course is what happens for quiver models with more fields. It has been shown for example \cite{Denef:2002ru,Denef:2007vg} that a quiver with three nodes and a closed loop has an exponential number of ground states (in the number of arrows). In addition, these theories exhibit an emergent $SL(2,\mathbb{R})$ symmetry on both their Higgs and Coulomb branches \cite{Anninos:2013nra,Anninos:2016szt}, meaning that they may be considered as toy models of black hole bound states. It has been argued in \cite{Bena:2012hf} that the large number of ground states are `pure-Higgs', in the sense that they have no support on the Coulomb branch. This seems to be in some tension with the notion that the ground states should morph smoothly as parameters are tuned, and that the wavefunction always has support in both the Higgs and Coulomb branch directions. One can envision studying a simple three-node quiver numerically in order to resolve this tension. Studying the $SL(2,\mathbb{R})$-invariant quantum mechanics on the Higgs and Coulomb branches reveals that the fields have distinct conformal dimensions on each branch \cite{Anninos:2013nra,Anninos:2016szt}. Thus, a simple three-node quiver may be regarded as a model whose study may provide intuition about the decoupling between the Higgs and Coulomb branches more generally. This intuition may be helpful in understanding the D1-D5 system \cite{Witten:1995zh,Witten:1997yu}.


\section*{Acknowledgements}
It is a pleasure to thank Dionysios Anninos, Arjun Bagchi, Nikolai Bobev, Frederik Denef, Daniel Z. Freedman, Jaehoon Lee, Hong Liu, Daniel Mayerson, Antonin Rovai, Edgar Shaghoulian, Shu-Heng Shao, Julian Sonner, Yifan Wang and Bert Vercnock for helpful discussions. Furthermore I am indebted to Matt Headrick's {\tt grassman.m} package. I am also indebted to Greg Moore for sharing some of his unpublished notes on the subject. I would like to especially thank Dionysios Anninos and Frederik Denef for stimulating discussions and comments on the draft. This work is supported by the U.S. Department of Energy under grant Contract Number  DE-SC0012567 and by NSF grant PHY-0967299.

\appendix

\section{Operators in spherical coordinates}\label{sphereapp}
It will be most convenient to work in spherical coordinates $(x,y,z)\rightarrow (r,\vartheta,\varphi)$ in which the matrix
\begin{equation}
\vec{\mathbf{x}}\cdot\vec{\boldsymbol{\sigma}}_{\alpha}^{~\,\beta} =r
\begin{pmatrix}
 \cos\vartheta &e^{-i\varphi} \sin\vartheta\\
e^{i\varphi} \sin\vartheta & - \cos\vartheta
\end{pmatrix}~.
\end{equation}
Furthermore we use the identity $\left(\vec{\mathbf{a}}\cdot\vec{\boldsymbol{\sigma}}_{\alpha}^{~\,\gamma}\right)\left(\vec{\mathbf{b}}\cdot\vec{\boldsymbol{\sigma}}_{\gamma}^{~\,\beta}\right)=\left(\vec{\mathbf{a}}\cdot\vec{\mathbf{b}}\right)\,\delta_\alpha^{~\,\beta}+i\left(\vec{\mathbf{a}}\times\vec{\mathbf{b}}\right)\cdot\vec{\boldsymbol{\sigma}}_{\alpha}^{~\,\beta} $ to write
\begin{align*}
\nabla_{\vec{\mathbf{x}}}\cdot\vec{\boldsymbol{\sigma}}_\alpha^{~\,\gamma}&=\left(\hat{\mathbf{r}}\cdot\vec{\boldsymbol{\sigma}}_\alpha^{~\,\omega}\right)\left(\hat{\mathbf{r}}\cdot\vec{\boldsymbol{\sigma}}_\omega^{~\,\eta}\right)\left(\nabla_{\vec{\mathbf{x}}}\cdot\vec{\boldsymbol{\sigma}}_\eta^{~\,\gamma}\right)\\
&=\left(\hat{\mathbf{r}}\cdot\vec{\boldsymbol{\sigma}}_\alpha^{~\,\omega}\right)\left[\delta_\omega^{~\,\gamma}\partial_r+\frac{i}{r}\left(\vec{\mathbf{x}}\times\nabla_{\vec{\mathbf{x}}}\right)\cdot\vec{\boldsymbol{\sigma}}_\omega^{~\,\gamma}\right]\\
&=\frac{1}{r}\left(\vec{\mathbf{x}}\cdot\vec{\boldsymbol{\sigma}}_\alpha^{~\,\omega}\right)\left[\delta_\omega^{~\,\gamma}\partial_r-\frac{1}{r}\,\vec{\mathbf{L}}\cdot\vec{\boldsymbol{\sigma}}_\omega^{~\,\gamma}\right]~.
\end{align*}
and the different components of the angular momentum are:
\begin{align}
\mathbf{L}^1&=i\left(\sin\varphi\,\partial_\vartheta+\cos\varphi\,\cot\vartheta\,\partial_\varphi\right)~,\\
\mathbf{L}^2&=-i(\cos\varphi\,\partial_\vartheta-\sin\varphi\,\cot\vartheta\,\partial_\varphi)~,\\
\mathbf{L}^3&=-i\,\partial_\varphi~.
\end{align}
The supercharge operators as well as the angular momentum and $R$-charge operators are now easy to put into \emph{Mathematica} using Matt Headrick's {\tt grassman.m} package. Annihilation operators for fermions are implemented as Grassman derivatives on the $\bar{\psi}$ and $\bar{\lambda}$. For completeness we also include the well-known expression for the Laplacian in spherical coordinates
\begin{equation}
\nabla_{\vec{\mathbf{x}}}^2=\frac{1}{r^2}\partial_r\,r^2\,\partial_r+\frac{1}{r^2\sin\vartheta}\partial_\vartheta\,\sin\vartheta\,\partial_\vartheta+\frac{1}{r^2\sin^2\vartheta}\partial_\varphi^2~.
\end{equation}

\section{Restricted Schr\"{o}dinger equation in each R-charge sector }\label{wavefunceqs}
In this appendix, we provide the differential equations one has to solve in order to obtain the energy eigenstates in each $R$-charge sector. To do so we will construct the $SU(2)_J$ highest weight states, which satisfy $\mathbf{J}^3\Psi=j\Psi$ and $\mathbf{J}^+\Psi=0$, where $\mathbf{J}^+\equiv\mathbf{J}^1+i\mathbf{J}^2$.  We can fill out the remaining states in the representation by acting with $\mathbf{J}^-\equiv\mathbf{J}^1-i\mathbf{J}^2$.

Without loss of generality, we can write down the full set of gauge invariant wavefunctions, labeled by their respective $R$-charges:
\begin{align}
\Psi_{-2}&=e^{i\gamma} A\, \bar{\psi}^1\bar{\psi}^2|0\rangle\\
\Psi_{-1}&=\left\{B_\alpha\,\bar{\psi}^\alpha+e^{i\gamma} C_\alpha\,\bar{\psi}^1\bar{\psi}^2\bar{\lambda}^\alpha\right\}|0\rangle\\
\Psi_0&=\left\{e^{-i\gamma}D+E_{\alpha\beta}\,\bar{\psi}^\alpha\bar{\lambda}^\beta+e^{i\gamma}F\,\bar{\psi}^1\bar{\psi}^2\bar{\lambda}^1\bar{\lambda}^2\right\}|0\rangle\\
\Psi_{+1}&=\left\{e^{-i\gamma}G_\alpha\,\bar{\lambda}^\alpha+H_\alpha\,\bar{\psi}^\alpha\bar{\lambda}^1\bar{\lambda}^2\right\}|0\rangle\\
\Psi_{+2}&=e^{-i\gamma}I\,\bar{\lambda}^1\bar{\lambda}^2|0\rangle
\end{align}
where $A,\dots, I$ are functions of $\left(\tilde{r},\vec{\mathbf{x}}\right)$ and $|0\rangle$ is the fermionic ground state annihilated by $\psi_\alpha$ and $\lambda_\alpha$. These wavefunctions span the full 16-dimensional fermionic Hilbert space. We now provide the Schr\"{o}dinger equations in order of difficulty.

In the hopes of being self-contained, we recall here certain definitions that appear throughout the main text:
\begin{equation}
\mathcal{D}^2\equiv \frac{1}{2\mu\,r^2}\,\partial_r\,r^2\,\partial_r+\frac{1}{2\,\tilde{r}}\,\partial_{\tilde{r}}\,\tilde{r}\,\partial_{\tilde{r}}~,
\end{equation}
and
\begin{equation}
V\equiv\frac{\left(\frac{\tilde{r}^2}{2}+\theta\right)^2+\mu\,r^2\,\tilde{r}^2}{2\mu}~.
\end{equation}
For the sake of keeping expressions compact, we define the operator $\hat{
\mathcal{H}}\equiv -\mathcal{D}^2+V$.

\subsection{R=$-$2}

The highest weight state in the $R=-2$ sector takes the form
\begin{equation}
\Psi_{-2}=e^{i(\gamma+j\,\varphi)}\sin^j\vartheta\, A(\tilde{r},r) \, \bar{\psi}^1\bar{\psi}^2|0\rangle~.
\end{equation}
The Schr\"{o}dinger equation $H\Psi_{-2}=\mathcal{E}\,\Psi_{-2}$ applied to this state then reduces to
\begin{equation}\label{hrm2}
\left(\hat{\mathcal{H}}+\frac{j(j+1)}{2\,\mu\,r^2}+\frac{1}{2\,\tilde{r}^2}\right)A=\mathcal{E}\,A~.
\end{equation}

\subsection{R=$+$2}
The highest weight state in the $R=+2$ sector takes the form
\begin{equation}
\Psi_{+2}=e^{-i(\gamma-j\,\varphi)}\sin^j\vartheta\, I(\tilde{r},r) \,\bar{\lambda}^1\bar{\lambda}^2~.
\end{equation}
The Schr\"{o}dinger equation $H\Psi_{+2}=\mathcal{E}\,\Psi_{+2}$ applied to this state then reduces to
\begin{equation}\label{hrp2}
\left(\hat{\mathcal{H}}+\frac{j(j+1)}{2\,\mu\,r^2}+\frac{1}{2\,\tilde{r}^2}\right)I=\mathcal{E}\,I~.
\end{equation}
This is the same equation as in the $R=-2$ case, and we see that their spectra coincide as a result of the discrete symmetry (\ref{discretesymm}).

\subsection{R=$-$1}
The highest weight state in the $R=-1$ sector takes takes the form
\begin{equation}
\Psi_{-1}=\left\{B_\alpha\,\bar{\psi}^\alpha+e^{i\gamma} C_\alpha\,\bar{\psi}^1\bar{\psi}^2\bar{\lambda}^\alpha\right\}|0\rangle~,
\end{equation}
with
\begin{equation}
B_\alpha=e^{i\left(j-\tfrac{1}{2}\right)\varphi}\sin^{j-\tfrac{1}{2}}\vartheta\begin{pmatrix}\tilde{B}_1\left(\tilde{r},r\right)+ \cos\vartheta\,\tilde{B}_2\left(\tilde{r},r\right)\\ e^{i\varphi} \, \sin \vartheta\,\tilde{B}_2\left(\tilde{r},r\right)\end{pmatrix}~,
\end{equation}
and
\begin{equation}
 C_\alpha=i\mu^{1/2}e^{i\left(j-\tfrac{1}{2}\right)\varphi}\sin^{j-\tfrac{1}{2}}\vartheta\begin{pmatrix}\tilde{C}_1\left(\tilde{r},r\right)+ \cos\vartheta\,\tilde{C}_2\left(\tilde{r},r\right)\\ e^{i\varphi} \sin \vartheta\, \tilde{C}_2\left(\tilde{r},r\right)\end{pmatrix}~
\end{equation}
where we have included the factor of $i\mu^{1/2}$ in $C_\alpha$ to ensure that the restricted Schr\"{o}dinger equation, represented as a matrix on the $B_\alpha$ and $C_\alpha$ is manifestly real and symmetric. Plugging $\Psi_{-1}$ into $H\Psi_{-1}=\mathcal{E}\Psi_{-1}$ gives rise to a set of coupled equations for  $\tilde{B}_\alpha$ and $\tilde{C}_\alpha$ which can be conveniently expressed in the following way
\begin{equation}\label{schrorm1}
\begin{pmatrix}\hat{\mathcal{H}} +\frac{4\,j^2-1}{8\,\mu\,r^2} &r &\frac{\tilde{r}}{\mu^{1/2}} &0\\ r &\hat{\mathcal{H}}+\frac{4j(j+2)+3}{8\,\mu\,r^2} &0 &\frac{\tilde{r}}{\mu^{1/2}}\\ \frac{\tilde{r}}{\mu^{1/2}} &0 &\hat{\mathcal{H}}+\frac{4\,j^2-1}{8\,\mu\,r^2}+\frac{1}{2\,\tilde{r}^2} &0\\0 &\frac{\tilde{r}}{\mu^{1/2}} &0 &\hat{\mathcal{H}}+\frac{4j(j+2)+3}{8\,\mu\,r^2}+\frac{1}{2\tilde{r}^2}\end{pmatrix}\begin{pmatrix}\tilde{B}_1\\ \tilde{B}_2\\ \tilde{C}_1\\ \tilde{C}_2\end{pmatrix}=\mathcal{E}\begin{pmatrix}\tilde{B}_1\\ \tilde{B}_2\\ \tilde{C}_1\\ \tilde{C}_2\end{pmatrix}~.
\end{equation}
It is worth noting that there are no $j=0$ solutions save for the trivial solution that satisfy $\mathbf{J}^-\Psi_{-1}=0$. Hence the states in this sector have $j\geq1/2$.

\subsection{R=$+$1}
The highest weight state satisfying $\mathbf{J}^3\Psi_{+1}=j\Psi_{+1}$ and $\mathbf{J}^+\Psi_{+1}=0$ in the $R=+1$ sector takes the form
\begin{equation}
\Psi_{+1}=\left\{e^{-i\gamma}G_\alpha\,\bar{\lambda}^\alpha+H_\alpha\,\bar{\psi}^\alpha\bar{\lambda}^1\bar{\lambda}^2\right\}|0\rangle
\end{equation}
with
\begin{equation}
\,G_\alpha= e^{i\left(j-\tfrac{1}{2}\right)\varphi}\sin^{j-\tfrac{1}{2}}\vartheta\begin{pmatrix}\tilde{G}_1\left(\tilde{r},r\right)+ \cos\vartheta\,\tilde{G}_2\left(\tilde{r},r\right)\\ e^{i\varphi} \, \sin \vartheta\,\tilde{G}_2\left(\tilde{r},r\right)\end{pmatrix}~,
\end{equation}
and similarly
\begin{equation}
 H_\alpha=i\,\mu^{1/2}\,e^{i\left(j-\tfrac{1}{2}\right)\varphi}\sin^{j-\tfrac{1}{2}}\vartheta\begin{pmatrix}\tilde{H}_1\left(\tilde{r},r\right)+ \cos\vartheta\,\tilde{H}_2\left(\tilde{r},r\right)\\ e^{i\varphi}  \sin \vartheta\,\tilde{H}_2\left(\tilde{r},r\right)\end{pmatrix}~.
\end{equation}
Where we have again introduced the factor of $i\,\mu^{1/2}$ in $H_\alpha$ to simplify the form of the restricted equations. Interestingly we find that the restricted Schr\"{o}dinger equation in this sector can be expressed as
\begin{equation}\label{totalhrp1}
\begin{pmatrix}\hat{\mathcal{H}} +\frac{4\,j^2-1}{8\,\mu\,r^2} &r &\frac{\tilde{r}}{\mu^{1/2}} &0\\ r &\hat{\mathcal{H}}+\frac{4j(j+2)+3}{8\,\mu\,r^2} &0 &\frac{\tilde{r}}{\mu^{1/2}}\\ \frac{\tilde{r}}{\mu^{1/2}} &0 &\hat{\mathcal{H}}+\frac{4\,j^2-1}{8\,\mu\,r^2}+\frac{1}{2\,\tilde{r}^2} &0\\0 &\frac{\tilde{r}}{\mu^{1/2}} &0 &\hat{\mathcal{H}}+\frac{4j(j+2)+3}{8\,\mu\,r^2}+\frac{1}{2\tilde{r}^2}\end{pmatrix}\begin{pmatrix}\tilde{H}_1\\ \tilde{H}_2\\ \tilde{G}_1\\ \tilde{G}_2\end{pmatrix}=\mathcal{E}\begin{pmatrix}\tilde{H}_1\\ \tilde{H}_2\\ \tilde{G}_1\\ \tilde{G}_2\end{pmatrix}~.
\end{equation}
which is the same as (\ref{schrorm1}), which again is a result of the symmetry (\ref{discretesymm}).

\subsection{R=0}\label{r0eqsapp}
For $\Psi_0$ a highest weight state, we repeat the expressions found in the main text, that is
\begin{equation}
\Psi_0=\left\{e^{-i\gamma}D+E_{\alpha\beta}\,\bar{\psi}^\alpha\bar{\lambda}^\beta+e^{i\gamma}F\,\bar{\psi}^1\bar{\psi}^2\bar{\lambda}^1\bar{\lambda}^2\right\}|0\rangle~,
\end{equation}
with
\begin{align}
D&=i\,e^{ij\varphi}\sin^j\vartheta\,\tilde{D}(\tilde{r},r)~,\quad\quad\quad F=i\,\mu^{1/2}\,e^{ij\varphi}\sin^j\vartheta\,\tilde{F}(\tilde{r},r)~,\\
E_{\alpha\beta}&=e^{ij\varphi}\sin^j\vartheta\,\begin{pmatrix}& e^{-i\varphi}\csc\vartheta\left[\tilde{E}_{11}+\cos\vartheta\left(\tilde{E}_{12}+\tilde{E}_{21}+\cos\vartheta\tilde{E}_{22}\right)\right] &\tilde{E}_{12}+\cos\vartheta\,\tilde{E}_{22}\\&\tilde{E}_{21}+\cos\vartheta\,\tilde{E}_{22} & e^{i\varphi}\sin\vartheta\,\tilde{E}_{22}\end{pmatrix}~,
\end{align}
Before writing down the Schr\"{o}dinger equation for the restricted set of functions, let us mention that $\tilde{D}$ satisfies the same differential equation as $\tilde{F}/\mu^{1/2}$. Unlike in the BPS sector, where the equations implied strict equality, we can not immediately come to the same conclusion here. In fact, away from zero energy it is not always possible to argue that the functions should be equal. The Schr\"{o}dinger equation implies that $\tilde{F}$ satisfies
\begin{equation}
\left(\hat{\mathcal{H}}-\mathcal{E}+\frac{j(j+1)}{2\,\mu\,r^2}+\frac{1}{2\,\tilde{r}^2}\right)\tilde{F}=\frac{\tilde{r}}{\mu^{1/2}}\left(\tilde{E}_{12}-\tilde{E}_{21}\right)
\end{equation}
and similarly for $\tilde{D}$. Let us assume that $\tilde{D}=\frac{1}{\mu^{1/2}}\left(\tilde{F}+\Phi(\tilde{r},r)\right)$~, then $\Phi$ satisfies
\begin{equation}
\left(\hat{\mathcal{H}}+\frac{j(j+1)}{2\,\mu\,r^2}+\frac{1}{2\,\tilde{r}^2}\right)\Phi=\mathcal{E}\Phi
\end{equation}
which is the restricted Schr\"{o}dinger equation of the $R=\pm 2$ sectors, which is expected to have non-zero solutions for certain $\mathcal{E}$ and $j$. 
We are now ready to write the coupled equations in the $R=0$ sector
\begin{equation}\label{totalHr0}
\begin{pmatrix}\hat{\mathcal{H}}+\frac{j(j+1)}{2\,\mu\,r^2}+\frac{1}{2\,\tilde{r}^2} &0 &0 &0 &0 &0 \\0 &\hat{\mathcal{H}} +\frac{j(j+1)}{2\,\mu\,r^2}+\frac{1}{2\tilde{r}^2} &0 &\frac{\tilde{r}}{\mu^{1/2}} &-\frac{\tilde{r}}{\mu^{1/2}} &0\\0 &0 &\hat{\mathcal{H}}+\frac{j(j-1)}{2\,\mu\,r^2} &0 &r &-\frac{1}{\mu\,r^2} \\\frac{\tilde{r}}{\mu^{1/2}}  &\frac{2\tilde{r}}{\mu^{1/2}} &0  &\hat{\mathcal{H}}+\frac{j(j+1)}{2\,\mu\,r^2} &0 &r\\-\frac{\tilde{r}}{\mu^{1/2}} &-\frac{2\tilde{r}}{\mu^{1/2}} &r &0 &\hat{\mathcal{H}}+\frac{j(j+1)}{2\,\mu\,r^2} & 0 \\0 &0 &0 &r &0 &\hat{\mathcal{H}}+\frac{(j+1)(j+2)}{2\,\mu\, r^2}\end{pmatrix}\begin{pmatrix}\Phi\\ \tilde{F}\\ \tilde{E}_{11}\\ \tilde{E}_{12}\\ \tilde{E}_{21} \\ \tilde{E}_{22}\end{pmatrix}=\mathcal{E}\begin{pmatrix}\Phi \\\tilde{F}\\ \tilde{E}_{11}\\ \tilde{E}_{12}\\ \tilde{E}_{21}\\ \tilde{E}_{22}\end{pmatrix}~.
\end{equation}
When solving these equations, we will encounter systems where $\Phi$ is nonzero. When this is the case the state $\Psi_0$ will come in a multiplet that includes states with $R$-charge $\pm 2$ with the same spin $j$. When $\Phi=0$, the multiplet containing $\Psi_0$ does not contain states with $R=\pm2$ and thus must be annihilated by the pair $Q_1\,Q_2$ and $\bar{Q}^1\,\bar{Q}^2$. When this happens we refer to it (in a slight abuse of terminology) as a short multiplet.

\section{Born-Oppenheimer approximation}\label{boapp}
Previous treatments of this model have assumed a separation of scales between the chiral multiplet degrees of freedom and those of the vector multiplet \cite{Denef:2002ru,Smilga:1986tv,Hori:2014tda}. This is natural given the D-brane interpretation of the degrees of freedom, where the chiral fields represent stretched strings at the intersection points between the branes. When the branes are well separated, the chiral fields become effectively heavy and can be integrated out. In practice, we can study the effective interactions of the remaining degrees of freedom in a Born-Oppenheimer approximation. This was explicitly done in \cite{Smilga:1986tv} and we repeat the analysis here.

The idea is to split the Hamiltonian $H$ and the supercharges in powers of $\mu^{-1/2}$ (recall from (\ref{ccrm}) that the gauginos come with an effective power of $\mu^{-1/2}$). Thus we can write $Q_\alpha=Q_\alpha^{(0)}+Q_\alpha^{(1)}$ with
\begin{align}
Q_\alpha^{(0)}&\equiv\sqrt{2}\left[\delta_\alpha^{~\,\gamma}\partial_{\bar{\phi}}-\phi\, \vec{\mathbf{x}}\cdot\vec{\boldsymbol{\sigma}}_{\alpha}^{~\,\gamma}\right]\left(\bar{\psi}\epsilon\right)_\gamma~,\\
Q_\alpha^{(1)}&\equiv i\left[\nabla_{\vec{\mathbf{x}}}\cdot\vec{\boldsymbol{\sigma}}_\alpha^{~\,\gamma}-\left(|\phi|^2+\theta\right)\delta_\alpha^{~\,\gamma}\right]\lambda_\gamma~,
\end{align}
and similarly $H=H^{(0)}+H^{(1)}+H^{(2)}$:
\begin{align}
H^{(0)}&\equiv-\partial_{\phi}\partial_{\bar{\phi}}+\vec{\mathbf{x}}^2\,|\phi|^2+\bar{\psi}\,\vec{\mathbf{x}}\cdot\vec{\boldsymbol{\sigma}}\,\psi~,\label{hamdecomp1}\\
H^{(1)}&\equiv \sqrt{2}\,i\left(\phi\,\left(\bar{\psi}\epsilon\right)\bar{\lambda}-\lambda\left(\epsilon\psi\right)\bar{\phi}\right)~,\label{hamdecomp2}\\
H^{(2)}&\equiv -\frac{1}{2\mu}\nabla_{\vec{\mathbf{x}}}^2+\frac{(|\phi|^2+\theta)^2}{2\mu}~.\label{hamdecomp3}
\end{align}
We assume that the ground state wavefunction of the system can be approximated as
\begin{equation}\label{bototwv}
\Psi^{\rm tot}= f_0(\vec{\mathbf{x}},\lambda)\Phi_0(\phi,\psi)+\sum_n f_n(\vec{\mathbf{x}},\lambda)\Phi_n(\phi,\psi)~,
\end{equation}
where $\Phi_0$ is the ground state of $H^{(0)}$, in which $\vec{\mathbf{x}}$ appears as a parameter, and the $\Phi_n$ are the excited states of $H^{(0)}$. It is easy to show that
\begin{equation}
\Phi_0=\sqrt{\frac{r}{\pi}}e^{-r|\phi|^2}w_\alpha\,\bar{\psi}^\alpha|0\rangle~,\quad\quad\quad w_\alpha=\begin{pmatrix}-e^{-i\varphi}\sin\frac{\vartheta}{2}\\\cos\frac{\vartheta}{2}\end{pmatrix}~,
\end{equation}
where we have expressed $\vec{\mathbf{x}}$ in spherical coordinates, and have normalized $\Phi_0$. The fermionic vacuum $|0\rangle$ is annihilated by $\psi_\alpha$ and $\lambda_\alpha$. There is some subtlety in the choice of $w_\alpha$, which we will come to shortly. Notice however that $H^{(0)}\,\Phi_0=G\,\Phi_0=0$.

We now must solve for $f_0$. In order to do so, we need to find the effective Hamiltonian which acts on it. Since this theory is supersymmetric, we can instead define the effective supercharges: $Q_\alpha^{\rm eff}=\left\langle Q_\alpha^{(1)}\right\rangle_{\Phi_0}+\dots$; these are
\begin{align}
Q_\alpha^{\rm eff}&=i\left[\left(\nabla_{\vec{\mathbf{x}}}-i\vec{\mathbf{A}}\right)\cdot\vec{\boldsymbol{\sigma}}_\alpha^{~\,\gamma}-\left(\frac{1}{2r}+\theta\right)\delta_\alpha^{~\,\gamma}\right]\lambda_\gamma~,\\
\bar{Q}^{\beta}_{\rm eff}&=i\,\bar{\lambda}^\gamma\left[\left(\nabla_{\vec{\mathbf{x}}}-i\vec{\mathbf{A}}\right)\cdot\vec{\boldsymbol{\sigma}}_\gamma^{~\,\beta}+\left(\frac{1}{2r}+\theta\right)\delta_\gamma^{~\,\beta}\right]~,
\end{align}
where
\begin{equation}
\vec{\mathbf{A}}=\frac{1-\cos\vartheta}{2\,r\,\sin\vartheta}\,\hat{\varphi}
\end{equation}
is the Berry's vector potential that arises from integrating out the chiral degrees of freedom. A different choice of $w_\alpha$ leads to a different gauge for this potential. Notice that $\vec{\mathbf{A}}$ is the vector potential of a Dirac monopole sitting at the origin, with a Dirac string along the negative $z$-axis. The effective supercharges satisfy an algebra\footnote{We used that $\nabla_{\vec{\mathbf{x}}}\times \vec{\mathbf{A}}=-\nabla_{\vec{\mathbf{x}}}\frac{1}{2r}$ in order to show this.} with non-vanishing commutator $\left\{Q_\alpha^{\rm eff},\bar{Q}^\beta_{\rm eff}\right\}=2\,\delta_\alpha^{~\,\beta}\,H^{\rm eff}$~, with
\begin{equation}\label{hameff}
H^{\rm eff}=-\frac{1}{2\mu}\left(\nabla_{\vec{\mathbf{x}}}-i\vec{\mathbf{A}}\right)^2+\frac{1}{2\mu}\left(\frac{1}{2r}+\theta\right)^2-\frac{\vec{\mathbf{x}}}{2\,r^3}\cdot\bar{\lambda}\,\vec{\boldsymbol{\sigma}}\,\lambda~.
\end{equation}
In summary, integrating out the chiral matter has generated an effective Hamiltonian which represents the behavior of a supersymmetric particle moving in the background of a dyon! D'Hoker and Vinet \cite{D'Hoker:1985et,D'Hoker:1985kb,Vinet:1985zh,D'Hoker:1986uh,D'Hoker:1987gr,D'Hoker:1990zy} studied this model in depth and algebraically determined its spectrum. Later Avery and Michelson \cite{Avery:2007xf} managed to write down the wavefunctions of this system for all energy levels.\footnote{This was possible because $H^{\rm eff}$ has an enhanced $SO(4)$ symmetry, and the states organize themselves into representations of this symmetry.}

To approximate the supersymmetric ground state, $f_0$ must be annihilated by $Q_{\alpha}^{\rm eff}$ and $\bar{Q}^\beta_{\rm eff}$. This yields
\begin{equation}
f_0= c\,\frac{e^{r\theta}}{\sqrt{r}}\,b_\alpha\,\bar{\lambda}^\alpha|0\rangle~,\quad\quad\quad b_\alpha=\begin{pmatrix}\cos\frac{\vartheta}{2}\\e^{i\varphi}\sin\frac{\vartheta}{2}\end{pmatrix}~,
\end{equation}
which is only normalizable for $\theta<0$, as expected, in which case $c=\sqrt{\frac{\mu\,\theta^2}{\pi}}$.

The wavefunction $\Psi^{\rm tot}$ in (\ref{bototwv}) can be viewed as an expansion of the exact BPS wavefunction in powers of $\mu^{-1/2}$ \cite{Mooreunpublished}.  We would like estimate how close the first order term $\chi_0\equiv f_0\Phi_0$ is to the actual ground state of the theory. Let us do so by computing the expectation value of the total Hamiltonian (\ref{ham}) in this state. Writing $\phi=\frac{\tilde{r}}{\sqrt{2}}e^{i\gamma}$ and integrating over the angular variables reveals
\begin{align}
\langle H\rangle_{\chi_0}&=\frac{4\theta^2}{\mu}\int_0^\infty d\tilde{r}\int_0^{\infty}dr\,\tilde{r}\,e^{-r\left(\tilde{r}^2-2\theta\right)}\left(1+r\,\tilde{r}^2-2r\,\theta\left(1-r\,\tilde{r}^2\right)\right)~.\nonumber\\
&\sim\int_0^{\infty}dr\,\frac{e^{2r\theta}}{r}~.
\end{align}
This quantity is logarithmically divergent at small $r$. The excited states of $H^{(0)}$, which have energies $\sim 2\,r$, become relevant near the origin and can not be neglected. 

Let us address some well warranted objections---after all, the Born-Oppenheimer approximation is typically phrased as a variational problem whereby one tries to minimize the functional $\langle H\rangle_{\chi_0}$. This is done in two steps, one first defines $\tilde{H}^{\rm eff}\equiv\langle H\rangle_{\Phi_0}$, with $\Phi_0$ given by the ground state of $H^{(0)}$, and then minimizes the functional $\left\langle \tilde{H}^{\rm eff}\right\rangle_{f_0}$ over $f_0$. Instead of following this procedure, we computed the effective supercharges, and used their algebra to find $H^{\rm eff}$ given in (\ref{hameff}). The only way for a discrepancy to arise is if $\tilde{H}^{\rm eff}\neq H^{\rm eff}$, which is indeed the case. Notice that $\left\langle H^{(1)}\right\rangle_{\Phi_0}=0$, so terms proportional to the gauginos in (\ref{hameff}) must come from higher order terms in perturbation theory, such as\cite{Smilga:1986rb,Smilga:2002mra}
\begin{equation}
\sum_{n>0}\frac{\left\lvert\left\langle\Phi_n|H^{(1)}|\Phi_0\right\rangle\right\rvert^2}{0-E_n}~.
\end{equation}
In conclusion, we have incorrectly dropped higher order terms in (\ref{bototwv}) whose contributions to the total energy (\ref{hameff}) come in at the same order in $\mu^{-1/2}$. This is nicely explained in \cite{Smilga:1986rb}. To obtain (\ref{hameff}) from the Born-Oppenheimer approximation without first computing the supercharges, one must include the higher order corrections in (\ref{bototwv}). This will lead to coupled equations of the form $f_n=\hat{L}_n\,f_0$, with $\hat{L}_n$ some set of differential operators. One then back-substitutes these expressions for the $f_n$ into the Schr\"{o}dinger equation for $f_0$ to find $H^{\rm eff}$ of (\ref{hameff}) acting on $f_0$. This indicates that the Born-Oppenheimer approximation of the supersymmetric ground state will include terms proportional to the excited states of $H^{(0)}$. These can be thought of as $D$ and $F$ in (\ref{r0wf}).


\end{document}